\def\year{2020}\relax
\documentclass[letterpaper]{article} 
\usepackage{aaai20}  
\usepackage{times}  
\usepackage{helvet} 
\usepackage{courier}  
\usepackage[hyphens]{url}  
\usepackage{graphicx} 

\usepackage{amsmath, amsthm}
\usepackage{amsfonts}       
\usepackage{algorithm}
\usepackage[noend]{algorithmic}
\usepackage{subcaption}
\usepackage{tabularx}
\usepackage{booktabs}

\newtheorem{theorem}{Theorem}

\newcommand\blfootnote[1]{%
	\begingroup
	\renewcommand\thefootnote{}\footnote{#1}%
	\addtocounter{footnote}{-1}%
	\endgroup
}

\usepackage{graphicx}  
\title{Understanding and Improving Proximity Graph based Maximum Inner Product Search}

\author{
	\Large 
	{Jie Liu*, 
		Xiao Yan*,
		Xinyan Dai,
		Zhirong Li,  
		James Cheng,
		Ming-Chang Yang}\\ 
	The Chinese University of Hong Kong\\
	\{jliu, xyan, xydai, zrli6, jcheng, mcyang\}@cse.cuhk.edu.hk
}

\begin{document}

\maketitle

\begin{abstract}
	
The inner-product navigable small world graph (ip-NSW) represents the state-of-the-art method for approximate maximum inner product search (MIPS) and it can achieve an order of magnitude speedup over the fastest baseline. However, to date it is still unclear where its exceptional performance comes from. In this paper, we show that there is a strong norm bias in the MIPS problem, which means that the large norm items are very likely to become the result of MIPS. Then we explain the good performance of ip-NSW as matching the norm bias of the MIPS problem --- large norm items have big in-degrees in the ip-NSW proximity graph and a walk on the graph spends the majority of computation on these items, thus effectively avoids unnecessary computation on small norm items. Furthermore, we propose the ip-NSW+ algorithm, which improves ip-NSW by introducing an additional angular proximity graph. Search is first conducted on the angular graph to find the angular neighbors of a query and then the MIPS neighbors of these angular neighbors are used to initialize the candidate pool for search on the inner-product proximity graph. Experiment results show that ip-NSW+ consistently and significantly outperforms ip-NSW and provides more robust performance under different data distributions. 
                	
\end{abstract}

\section{Introduction}\label{sec:intro}

\blfootnote{*Co-first authors are ranked alphabetically. Correspondence to Xiao Yan.}

For a query $q$, maximum inner product search (MIPS) finds an item that maximizes $q^{\top}x_i$ in a dataset $\mathcal{X}\!=\!\{x_i\in\mathbb{R}^d|i\!=\!1,\cdots,n\}$ containing $n$ items~\cite{conetree}. MIPS has a number of applications in recommender systems, computer vision and machine learning. Examples include recommendation based on user and item embeddings learned via matrix factorization~\cite{koren:mf}, object matching with visual descriptor~\cite{felzens:obj}, memory network training~\cite{chandar:mipsmemory} and reinforcement learning~\cite{jun:mipsreinforcement}. In practice, it is usually required to find the top-$k$ items having the largest inner product with $q$. When the dataset is large and the dimension (i.e., $d$) is high, exact MIPS is usually too costly and finding approximate MIPS (i.e., items with inner product close to the maximum) suffices for most applications. Therefore, we focus on approximate MIPS in this paper.

\textbf{Related work.} Due to its broad range of applications, many algorithms for MIPS have been proposed. Tree-based methods such as cone tree~\cite{conetree} and PCA tree~\cite{pcatree} were first used but they suffer from poor performance on high dimensional datasets. Locality sensitive hashing (LSH) based methods are widely used for similarity search~\cite{li2018general}. ALSH~\cite{shrivastava:alsh}, Simple-LSH~\cite{neyshabur:simple-lsh} and Norm-Range LSH~\cite{yan:normrange}, transform MIPS into Euclidean or angular similarity search and reuse existing hash functions. LEMP~\cite{teflioudi:lemp} and FEXIPRO~\cite{li:fexipro} target exact MIPS and adopt various pruning rules to avoid unnecessary computation. Maximus~\cite{MAXIMUS} shows that the pruning-based methods do not always outperform brute-force linear scans using optimized computation libraries.

\textbf{ip-NSW.} In a proximity graph, each item is connected to some items that are most similar to it w.r.t. a given similarity function~\cite{knngraph}. A similarity search query is processed by a walk in the graph, which keeps moving towards items that are most similar to the query. Proximity graph based methods achieve excellent recall-time performance\footnote{Recall-time performance measures the time taken to reach a given recall for query processing.} for Euclidean distance nearest neighbor search (Euclidean NNS) and an number of variants have been proposed~\cite{spreading-out,harwood2016fanng,bridge}. Among them, the navigable small word graph (NSW)~\cite{malkov2014nsw} and its hierarchical version (HNSW)~\cite{HNSW} represent the state-of-the-art and we introduce NSW in greater details in Section~\ref{sec:ip-NSW}. Morozov and Babenko~(\citeyear{morozov:graphmips}) showed that NSW also works well for MIPS. They proposed the ip-NSW algorithm, which directly uses inner product as similarity function to construct and search NSW. ip-NSW outperforms all existing MIPS algorithms (including those mentioned in the related work) by a large margin in terms of recall-time performance and the speedup can be an order of magnitude for achieving the same recall~\cite{morozov:graphmips}. 

In spite of its excellent performance, there lacks a good understanding why ip-NSW works well for MIPS. Morozov and Babenko~(\citeyear{morozov:graphmips}) proved that a greedy walk in the proximity graph will find the exact MIPS of a query if the graph is the Delaunay graph for inner product. Nevertheless, the ip-NSW graph is only an approximation of the Delaunay graph, which contains much more edges than the ip-NSW graph. It is not clear how accurately the ip-NSW graph approximates the Delaunay graph and how the quality of the approximation affects the performance of ip-NSW. Moreover, their theory does not provide insights on how to improve the performance of ip-NSW. For proximity graph based similarity search algorithms, a rigorous theoretical justification is usually difficult due to the complexity of real datasets. In this case, an intuitive explanation is helpful if it leads to a better understanding of the algorithm and provides insights for performance improvements.

\textbf{Contributions.} We make three main contributions in this paper. Firstly, we identify an important property of the MIPS problem --- strong norm bias, which means large norm items are much more likely to be the result of MIPS. Although it is common sense that MIPS is biased towards large norm items, the interesting thing is the intensity of the norm bias we observed. In the four datasets we experimented, items ranking top 5\% in norm occupy at least 87.5\% and as high as 100\% of the top-10 MIPS result. We also found that a skewed norm distribution, in which some items have much larger norm than others, is not a must for the strong norm bias to appear, and the large cardinality of modern datasets is also an important reason behind the strong norm bias.

Secondly, we explain the excellent performance of ip-NSW as matching the norm bias of the MIPS problem. We found that items with large norm have much higher in-degree than the average in the proximity graph built by ip-NSW and a graph walk spends a dominant portion of its computation on these items. Therefore, ip-NSW performs well for MIPS because it effectively avoids unnecessary computation on small-norm items, which are unlikely to be the results of MIPS.

Thirdly and most importantly, we propose the ip-NSW+ algorithm, which significantly improves the performance of ip-NSW. We found that the norm bias in ip-NSW can harm the performance of MIPS by spending computation on many large norm items that do not have a good inner product with the query. To tackle this problem, we introduce an additional angular proximity graph in ip-NSW+ and utilize the fact that items pointing to similar direction are likely to share similar MIPS neighbors. By retrieving the MIPS neighbors of the angular neighbors of the query, ip-NSW+ avoids computation on both small norm items and large norm items that do not have a good inner product with the query. To our knowledge, ip-NSW+ is the first similarity search algorithm that uses two proximity graphs constructed from different similarity functions. Experimental results show that ip-NSW+ not only significantly outperforms ip-NSW but also provides more robust performance under different data distributions.            

\section{Norm Bias in MIPS}\label{sec:norm bias}

\begin{table}[t]
	\centering
	\caption{Dataset statistics}
	\label{tab:datasets}
	\begin{center}
		\begin{sc}
			\fontsize{8}{9}\selectfont
			\begin{tabular}{cccl}
				\toprule		
				Dataset & \# items & \# dimensions \\ 
				\midrule
				Yahoo!Music & 136,736 & 300 \\
				WordVector & 1,000,000 & 300 \\
				ImageNet   & 2,340,373 & 150 \\
				Tiny5M    & 5,000,000 & 384  \\
				\bottomrule
				\vspace{-8mm}
			\end{tabular}
		\end{sc}
	\end{center}
\end{table}

In this section, we show that there exists strong norm bias in the MIPS problem. We also argue that large dataset cardinality also contributes to the norm bias.

\begin{figure*}[htb]
	\centering
	\includegraphics[width=0.24\textwidth]{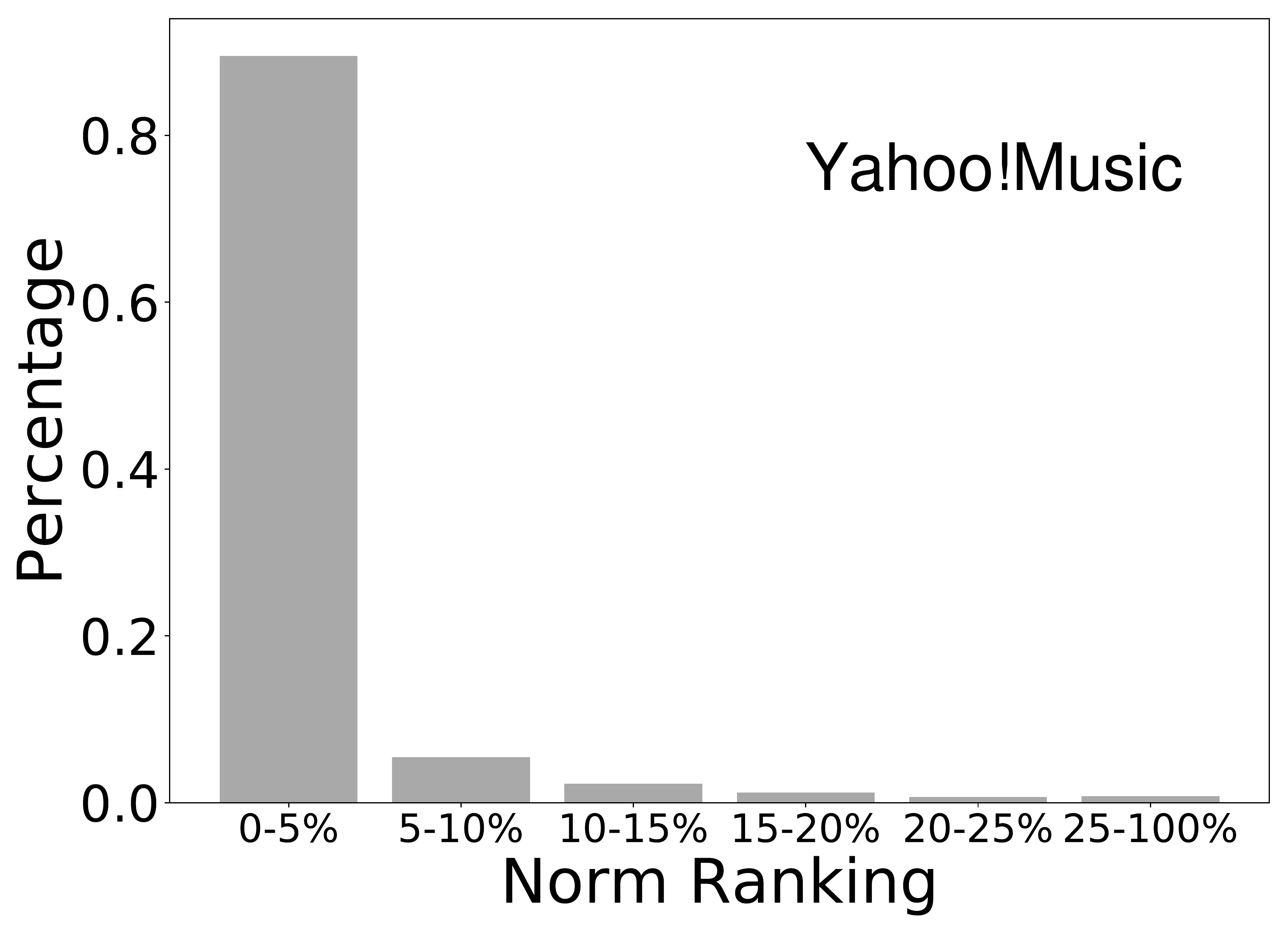}
	\includegraphics[width=0.24\textwidth]{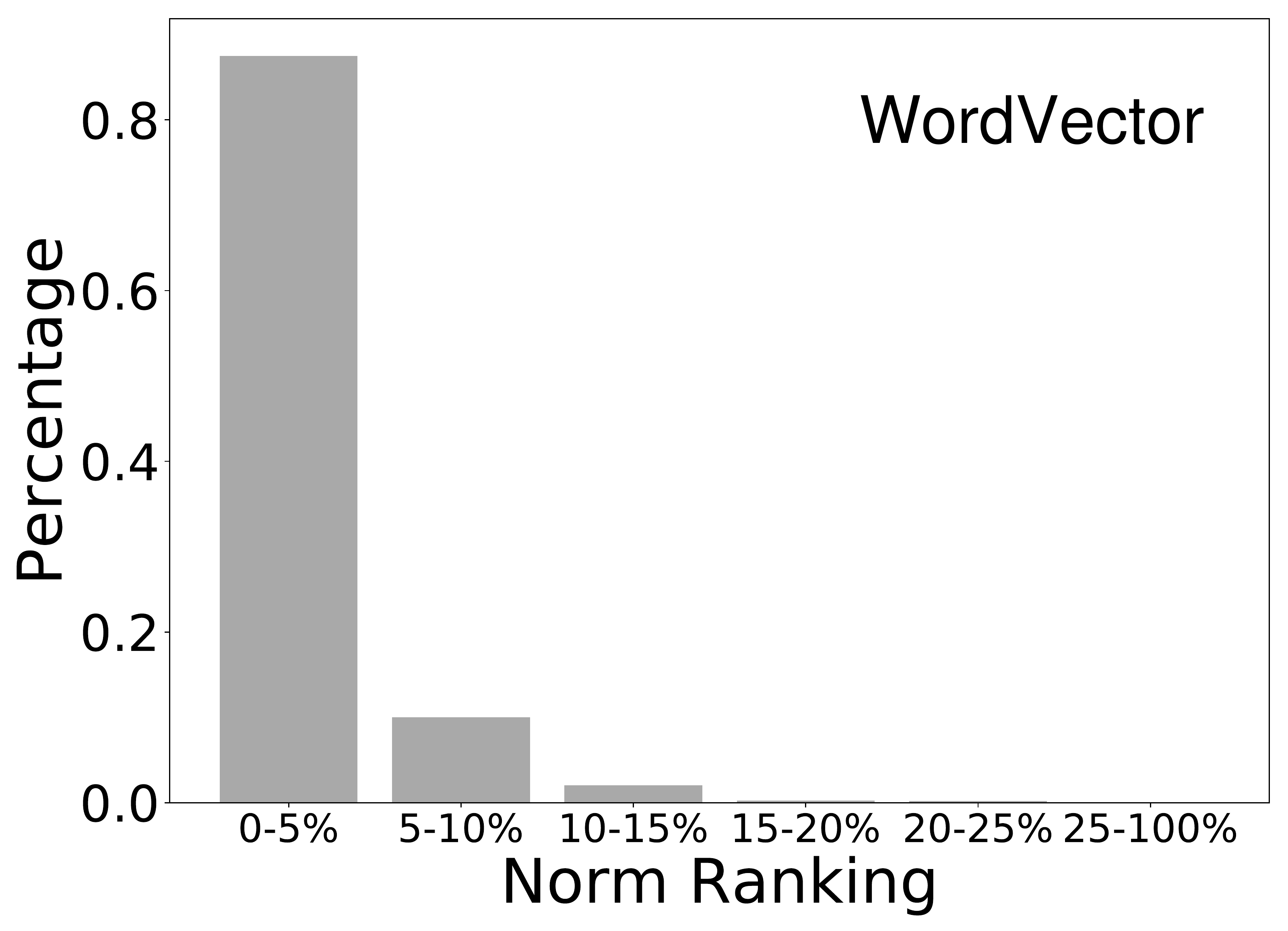}
	\includegraphics[width=0.24\textwidth]{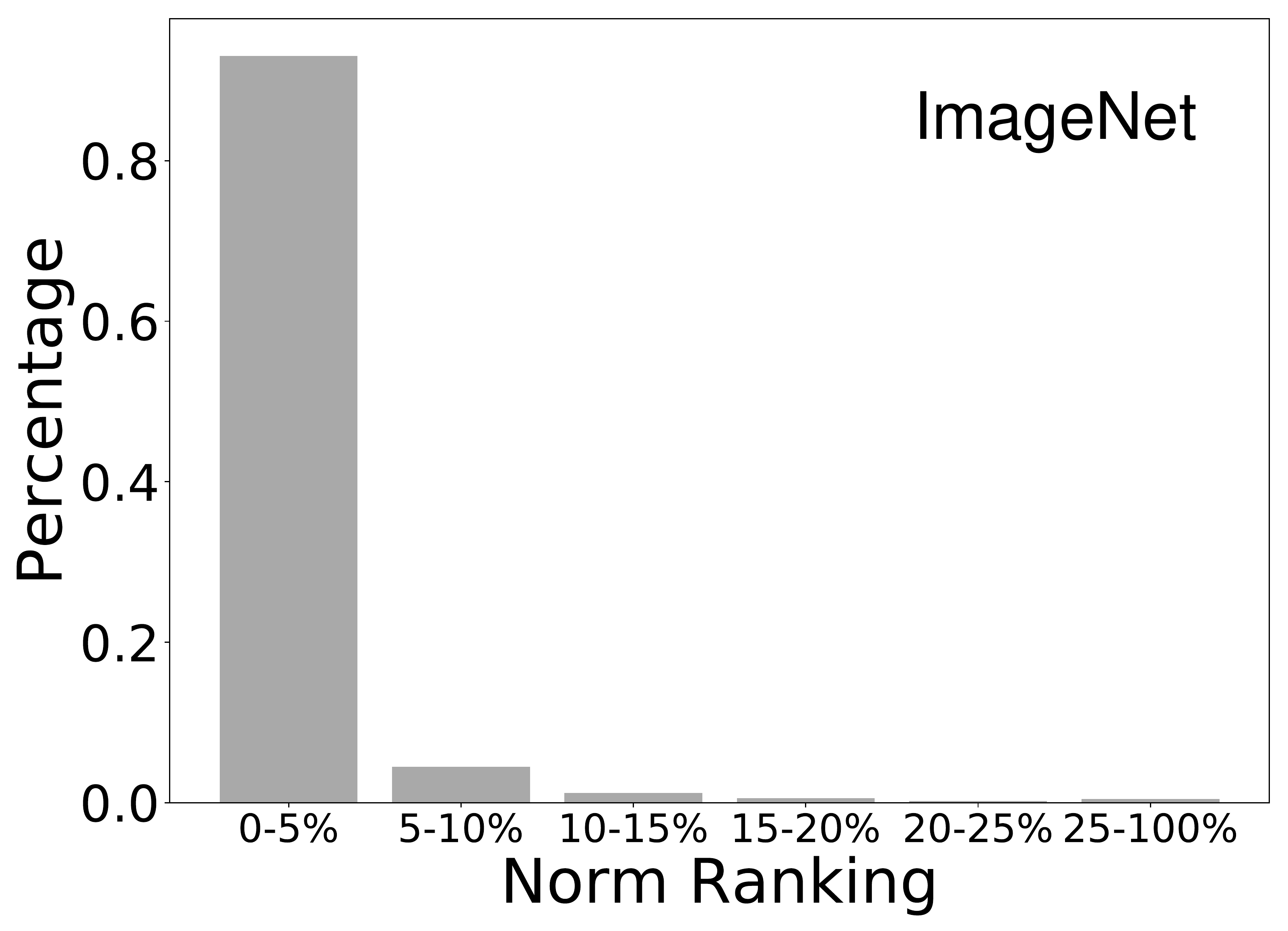}
	\includegraphics[width=0.24\textwidth]{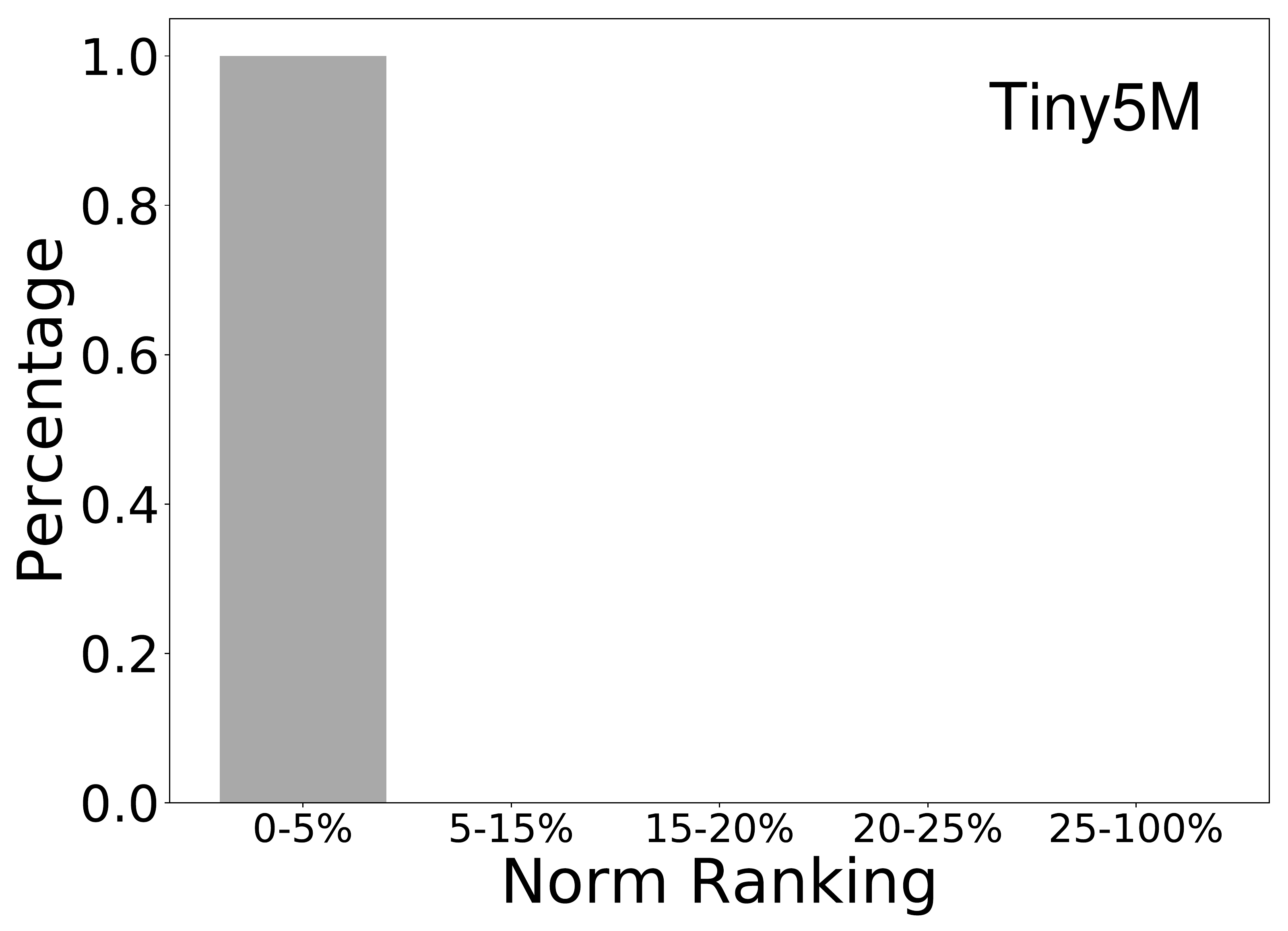}
	\caption{The percentage that items in each norm group occupy in the result set}
	\label{fig:result ranking}
\end{figure*}

To find out to what extent norm affects an item's chance of being the result of MIPS, we conducted the following experiment. We used four datasets, i.e., Yahoo!Music, WordVector, ImageNet and Tiny5M. Some statistics of the datasets can be found in Table~\ref{tab:datasets} and more details are given in Section 5. For each dataset, we found the exact top-10 MIPS~\footnote{Choosing top-10 MIPS is not arbitrary as it is widely adopted in related works.} result of 1,000 randomly selected queries using linear scan, which gave us a result set containing 10,000 items (duplicate items exist as an item can be in the results of multiple queries). We also partitioned the items into groups according to their norm, e.g., items ranking top 5\% in norm and items ranking top 20\%-25\% in norm. Finally, for items in each norm group, we calculated the percentage they occupy in the result set, which is plotted in Figure~\ref{fig:result ranking}.

\begin{figure*}[htb]
	\centering
	\includegraphics[width=0.24\textwidth]{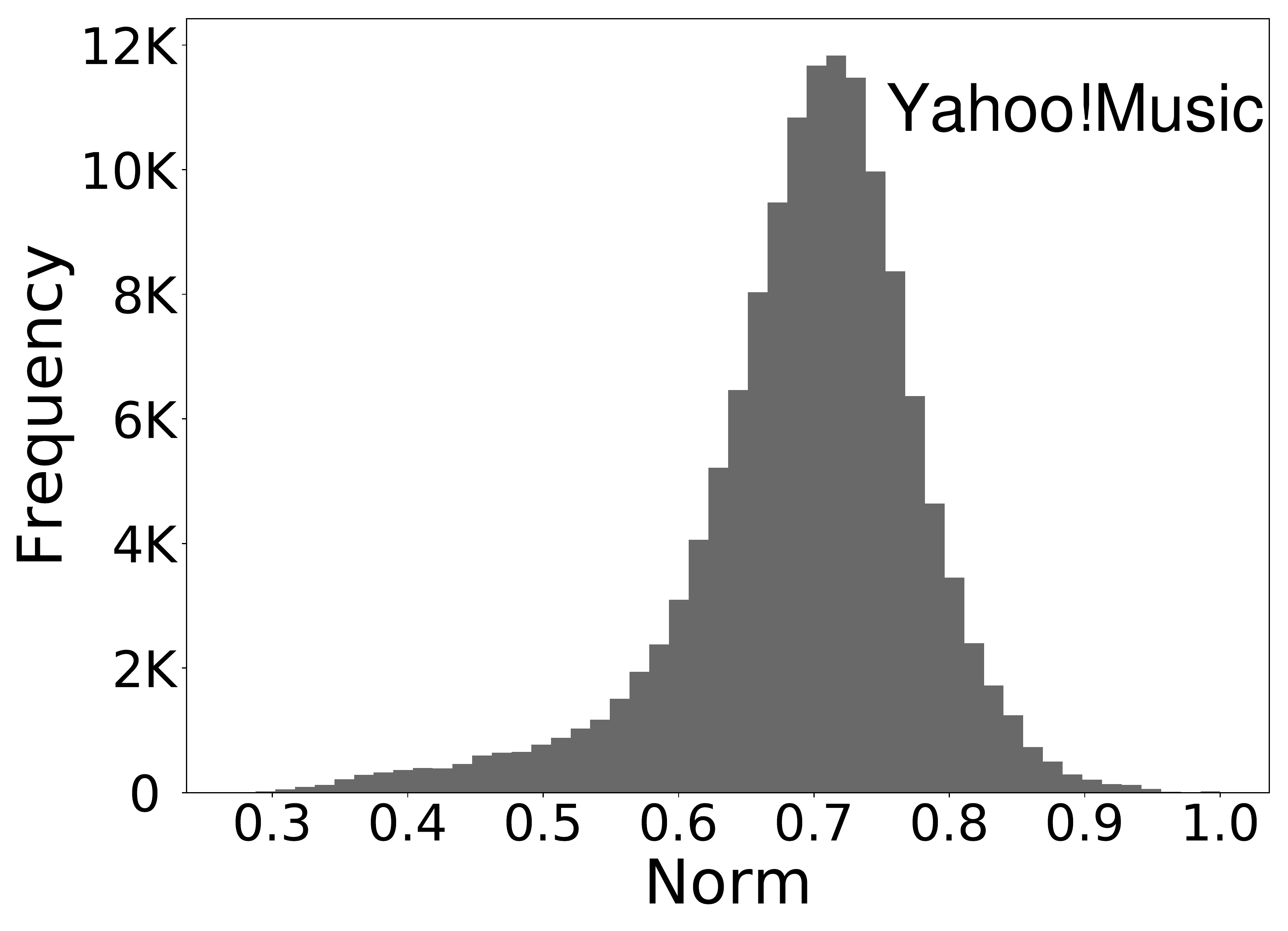}
	\includegraphics[width=0.24\textwidth]{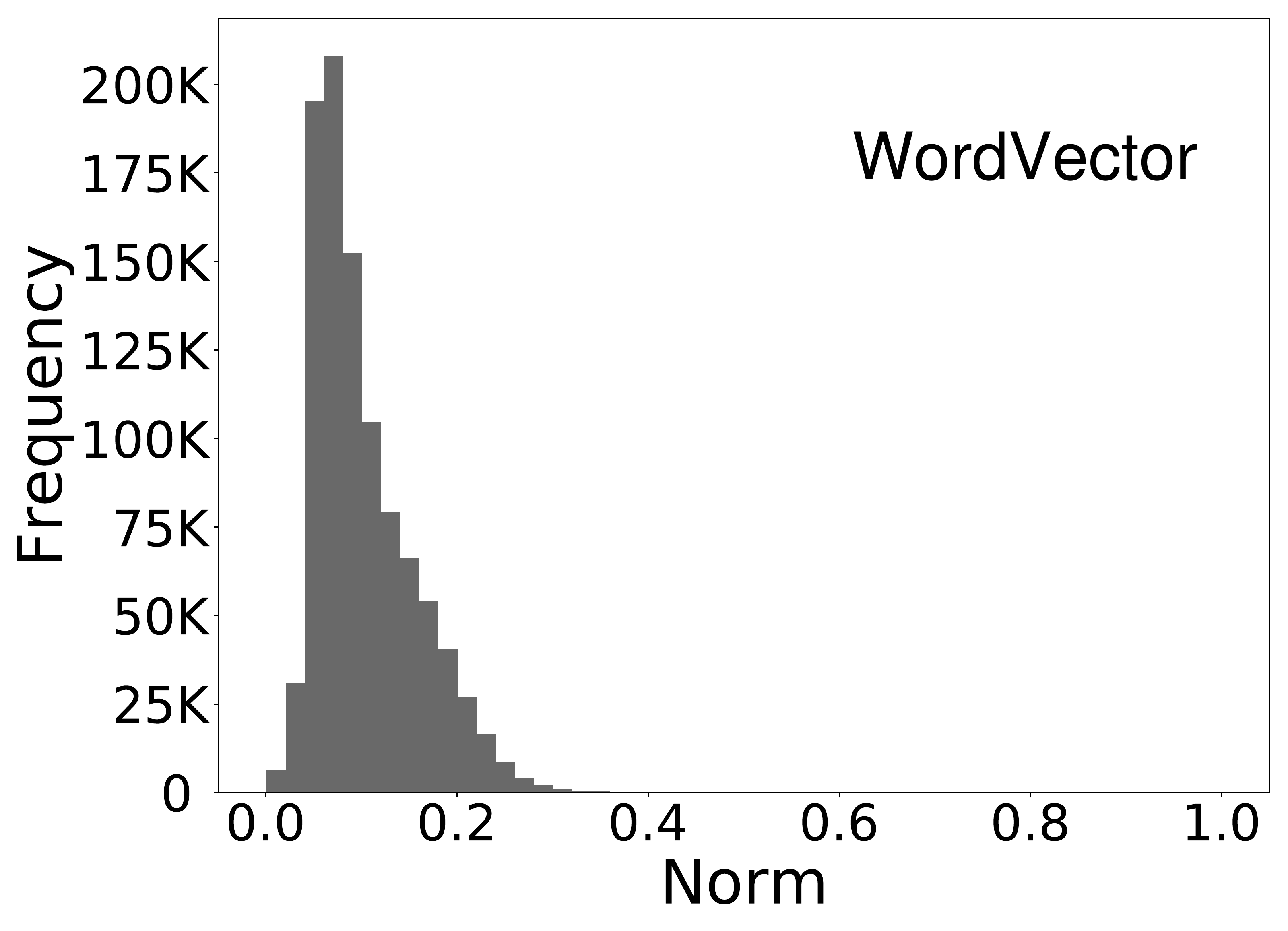}
	\includegraphics[width=0.24\textwidth]{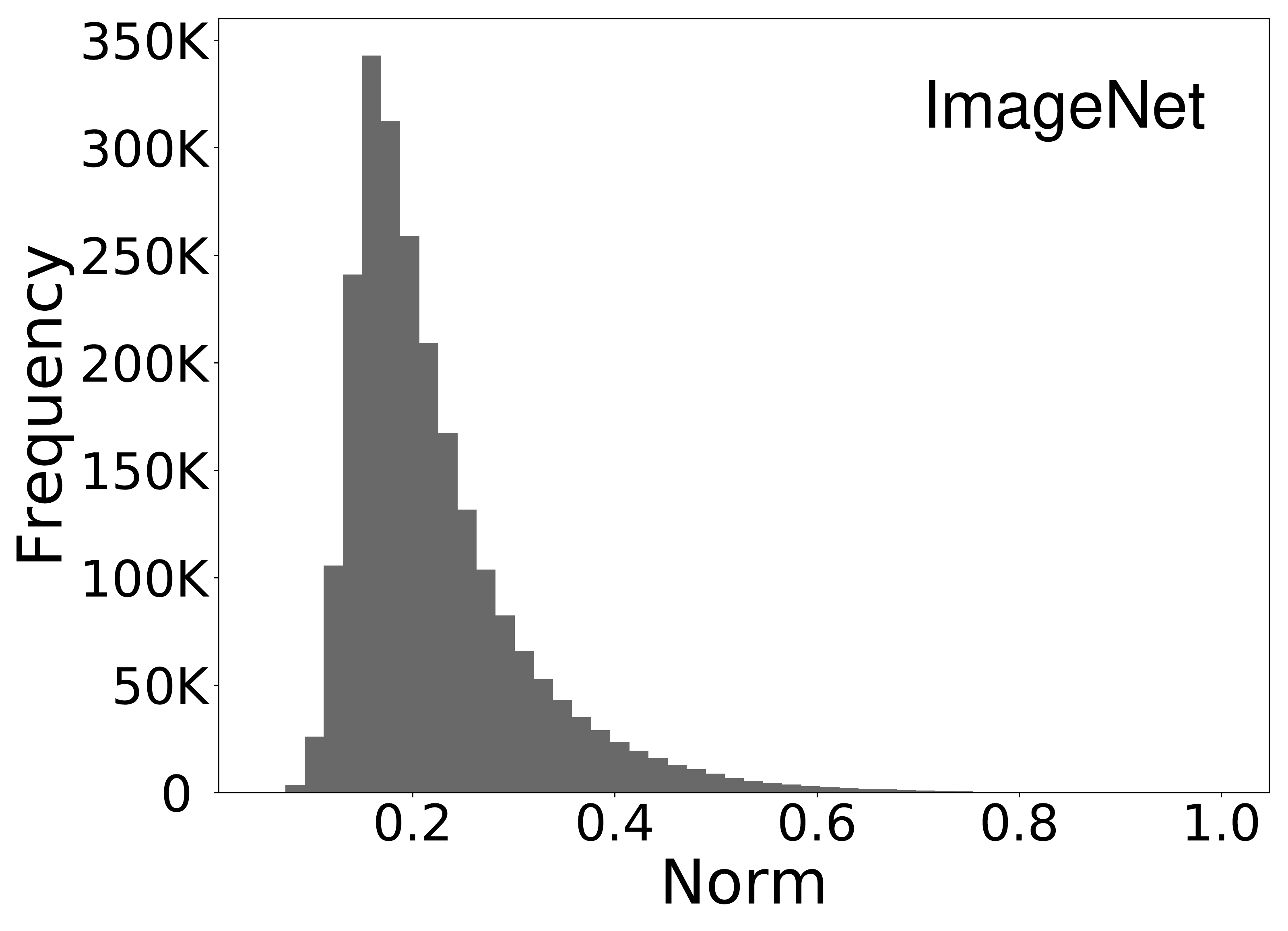}
	\includegraphics[width=0.24\textwidth]{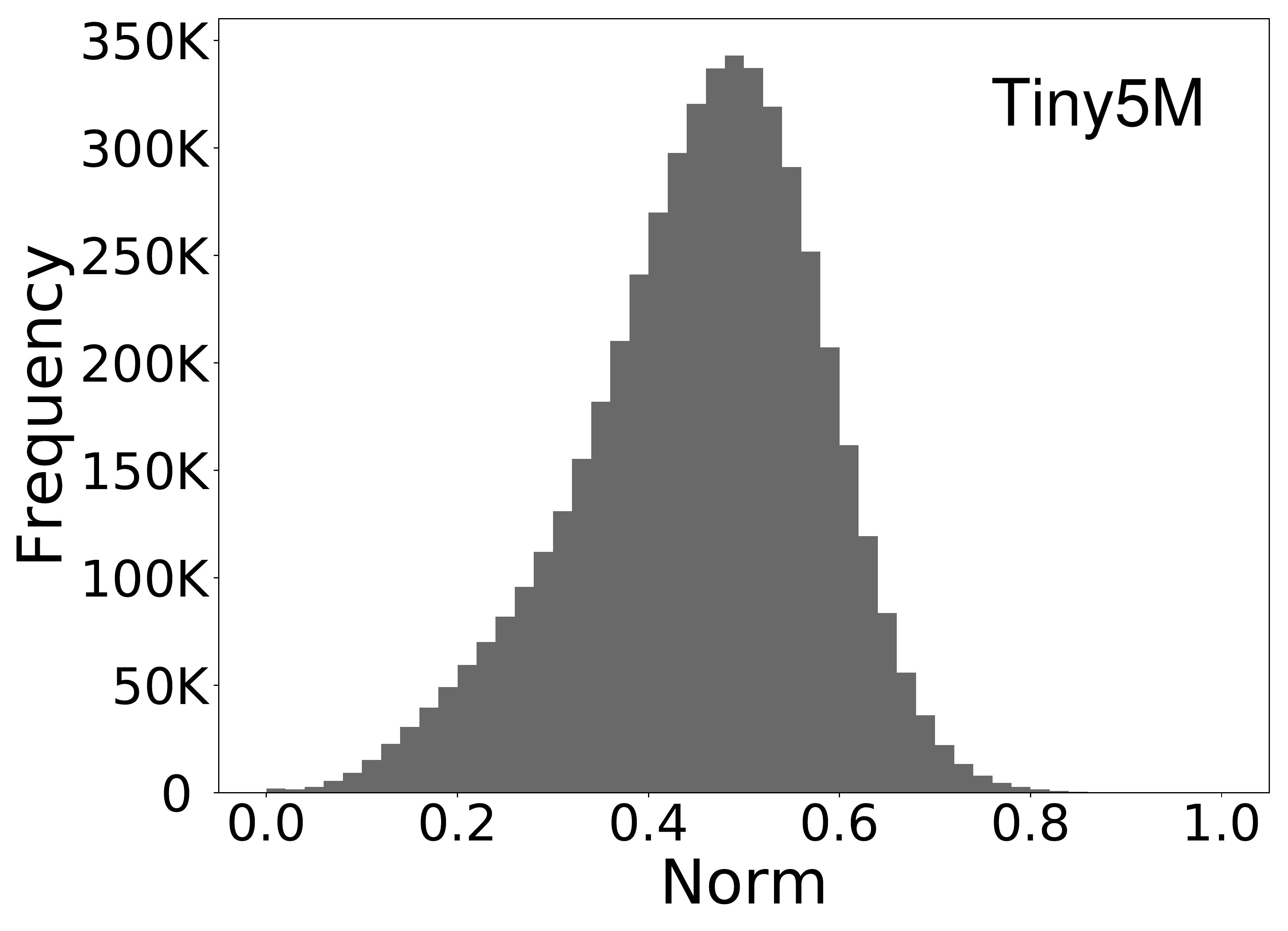}
	\caption{Norm distributions (maximum norm normalized to 1)}
	\label{fig:norm distribution}
\end{figure*}

Figure~\ref{fig:result ranking} shows that items with large norm are much more likely to be the result of MIPS. Specifically, items ranking top 5\% in norm take up 89.5\%, 87.5\%, 93.1\% and 100\% in the ground truth top-10 MIPS results for Yahoo!Music, WordVector, ImageNet and Tiny5M, respectively. One may conjecture that the norm bias is caused by skewed norm distribution, in which the top ranking items have much larger norm than the others. We plot the norm distribution of the datasets in Figure~\ref{fig:norm distribution} and it shows that this conjecture does not hold for Yahoo!Music and Tiny5M, in which most items have a norm close to the maximum. In fact, the 95\% percentile\footnote{We define $\eta_t$, the $t$\% percentile of the norm distribution, as $t=\frac{|\{x\in\mathcal{X}, \Vert x \Vert\le \eta_t\}|}{|\mathcal{X}|}\times 100$.} of the norm distribution is only 1.16 times of the median norm for Yahoo!Music (1.37 for Tiny5M). Theorem~\ref{theorem:bias} also shows that skewed norm distribution alone is not enough to explain the strong norm bias we observed.

\begin{theorem}\label{theorem:bias}
	For two independent random vectors $x$ and $y$ in $\mathbb{R}^d$, the entries of $x$ are independent and $x_i\!\sim\!\mathcal{N}(0,\alpha)$ for $i\!=\!1,2,\cdots,d$ with $\alpha \!\ge\!1$, the entries of $y$ are also independent and $y_i\!\sim\!\mathcal{N}(0,1)$ for $i\!=\!1,2,\cdots,d$. For a query $q\in\mathbb{R}^d$, we have $\mathbb{P}\lbrack q{\top}x\ge q{\top}y \mid  q{\top}x\ge 0, q{\top}y\ge 0 \rbrack = \frac{2}{\sqrt{\pi^2\alpha}} \int_{0}^{+\infty} e^{-\frac{a^2}{2\alpha}} \int_{0}^{a} e^{-\frac{b^2}{2}} \mathrm{d} b \mathrm{d} a$.  
\end{theorem}    

\begin{figure}[!t]	
	\centering 
	\begin{minipage}[b]{0.22\textwidth}
		\centering
		\includegraphics[width=\textwidth]{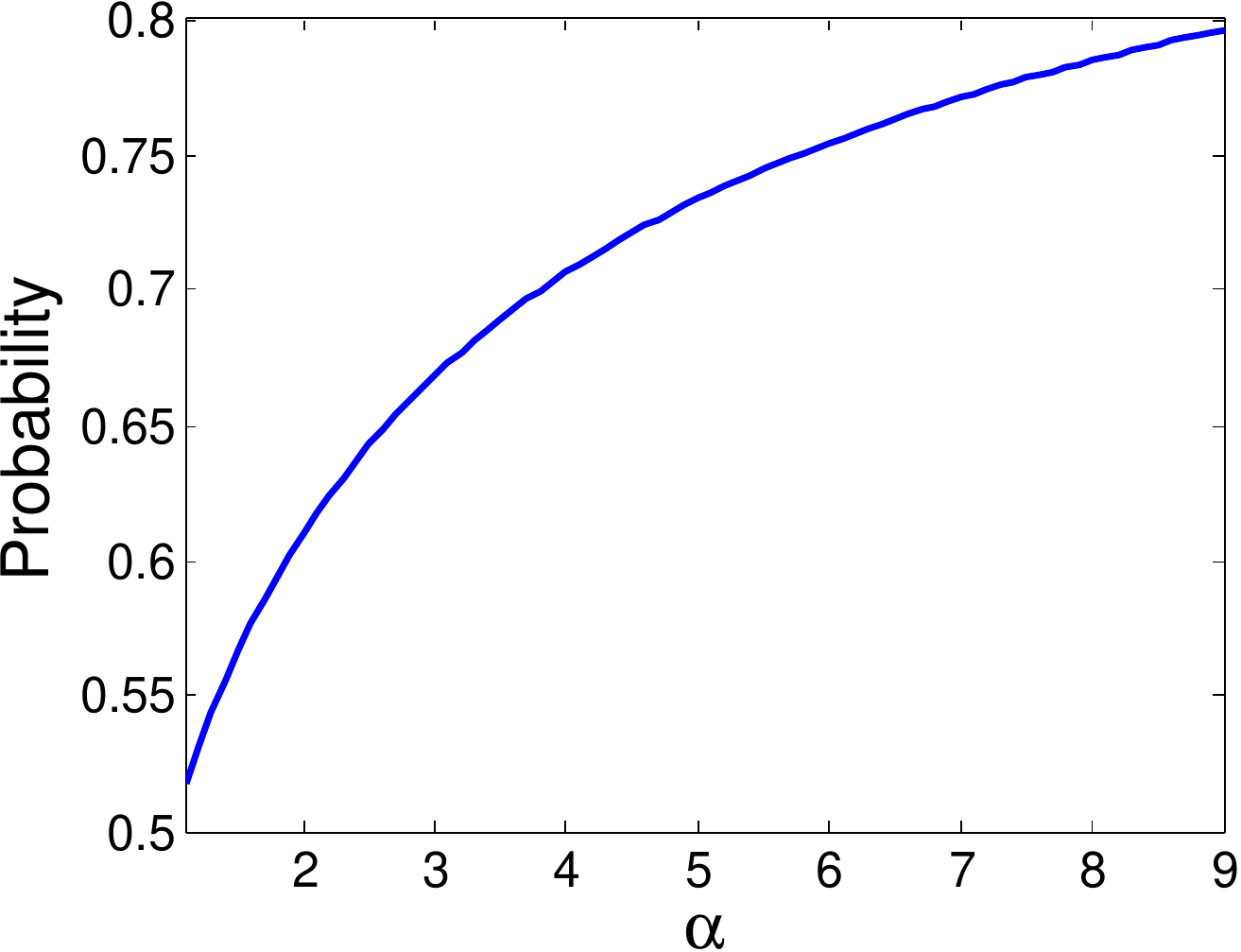}
		\subcaption{Norm bias vs. $\alpha$}\label{fig:norm and probability}
	\end{minipage}
	\begin{minipage}[b]{0.24\textwidth}
		\centering
		\includegraphics[width=\textwidth]{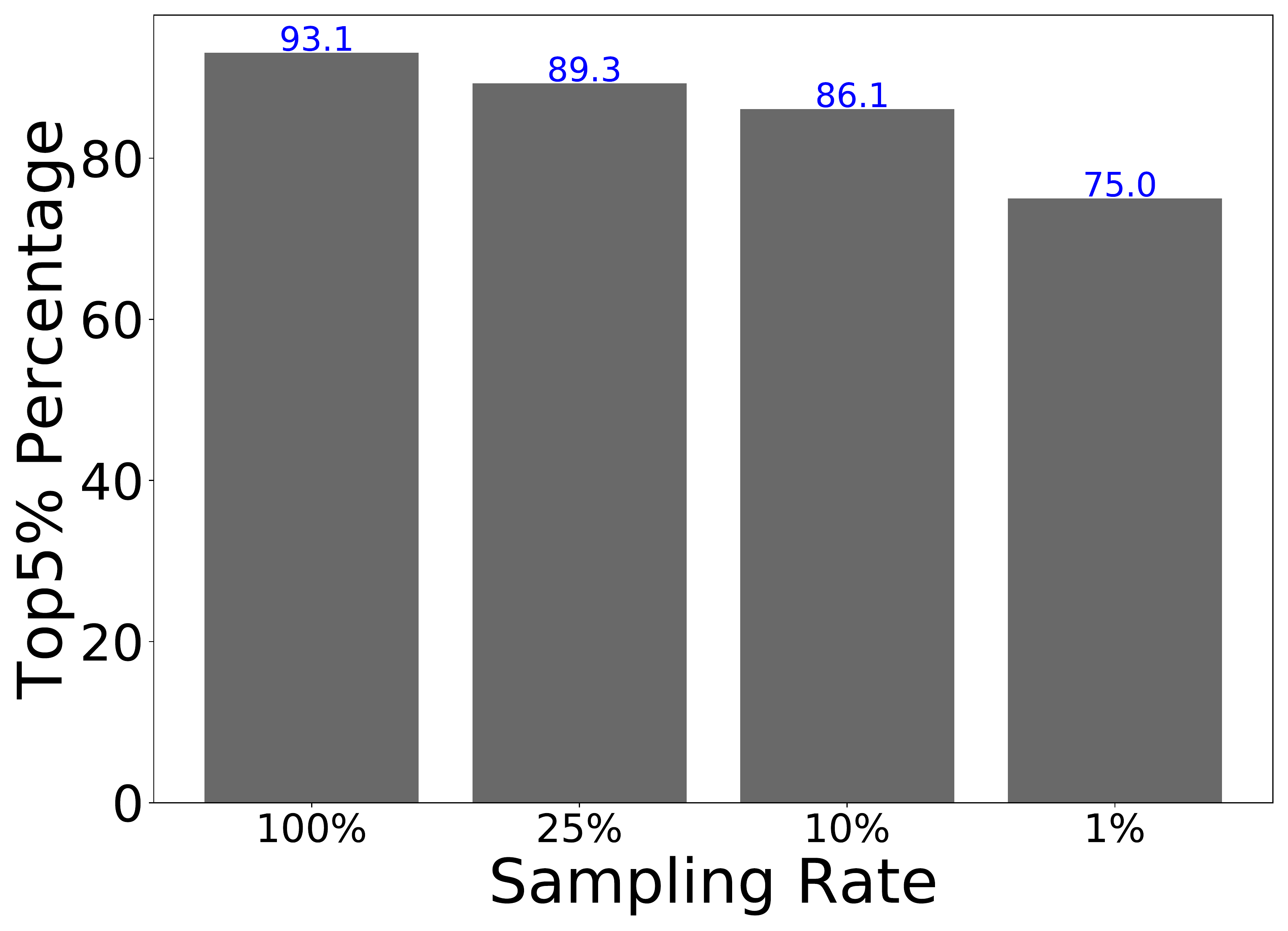}
		\subcaption{Norm bias vs. candinality}\label{fig:norm and cadinality}
	\end{minipage}
	\caption{Analysis of the norm bias}
	\label{fig:norm bias}
\end{figure}

The proof can be found in the supplementary material~\footnote{See https://arxiv.org/pdf/1909.13459.pdf for the supplementary material}. Intuitively, Theorem~\ref{theorem:bias} quantifies how likely larger norm will result in larger inner product. As $\mathbb{E}\lbrack \Vert x \Vert^2 \rbrack \!\!=\!\! \alpha d$ and $\mathbb{E}\lbrack \Vert y \Vert^2 \rbrack \!=\! d$, the norm of $x$ is roughly $\sqrt{\alpha}$ times of $y$. We constrain the inner products to be non-negative because negative inner product is not interesting for many practical applications such as recommendation. $\mathbb{P}\lbrack q{\top}x\ge q{\top}y \mid  q{\top}x\ge 0, q{\top}y\ge 0 \rbrack$ is a function of $\alpha$ and we plot its curve in Figure~\ref{fig:norm and probability} using numerical integration. The results show that larger norm only brings a modest probability (comparing with 0.5) of having larger inner product. For example, the probability of having larger inner product is only 0.56 with $\alpha\!=\!1.35$. Recall that the 95\% percentile norm is 1.16 times of the median for Yahoo!Music and $\sqrt{1.35} \approx1.16$. However, the observed norm bias (items ranking top 5\% in norm take up 89.5\% of the top-10 MIPS result for Yahoo!Music) is much stronger than that is predicted by the norm distribution and this is also true for WordVector, ImageNet and Tiny5M.

We find that large dataset cardinality also contributes to the norm bias. Consider an item $x$ with modest norm and there are $m$ items having larger norm than $x$ in the dataset. Item $x$ only has a probability of $p=\prod_{i=1}^{m}p_i$ to be the MIPS of a query (if we assume all items are independent), in which $p_i=\mathbb{P}\lbrack q{\top}x\ge q{\top}z^i \mid  q{\top}x\ge 0, q{\top}z^i\ge 0 \rbrack$ and $z^i$ is the $i$-th item that has larger norm than $x$. As $p_i<0.5$ and $m$ is large for large datasets, the probability $p$ is very small. This explanation suggests that the norm bias is stronger for larger datasets even if the norm distribution is the same. To validate, we uniformly sample the ImageNet dataset and plot the percentage that items ranking top 5\% in norm occupy in the top-10 MIPS result in Figure~\ref{fig:norm and cadinality}. Note that uniform sampling ensures that the shape of the norm distribution is the same across different sampling rate but a lower sampling rate results in smaller dataset cardinality. The results show that the top norm items take up a greater portion of the MIPS results under larger dataset cardinality, which validates our analysis. Our explanation justifies the extremely strong norm bias observed on the Tiny5M dataset even if its norm distribution is not skewed. Moreover, this explanation also implies that strong norm bias may be a universal phenomenon for modern datasets as they usually have large cardinality. 
               
\section{Understanding the Performance of ip-NSW}\label{sec:ip-NSW}

In this section, we briefly introduce the ip-NSW algorithm and show that ip-NSW has excellent performance because it matches the strong norm bias of the MIPS problem.

\subsection{NSW}

The query processing and index construction procedures of NSW are shown in Algorithm~\ref{alg:nsw query} and Algorithm~\ref{alg:nsw graph construction}, respectively. In Algorithm~\ref{alg:nsw query}, a graph walk for a similarity search query $q$ starts at an entry vertex $v_0$ (chosen randomly or deterministically) and keeps probing the neighbors of the unchecked vertex that is most similar to $q$ in the candidate pool $\mathcal{C}$. The size of the candidate pool, $l$, controls the quality of the search results and the graph walk is more likely to get stuck at local optimum under small $l$~\footnote{A graph walk with $l=1$ is usually called greedy search.}. 

For index construction, NSW does not require each item to connect to its exact top-$M$ neighbors in the dataset. Items are inserted sequentially into the graph in Algorithm~\ref{alg:nsw graph construction} and Algorithm~\ref{alg:nsw query} is used to find the approximate top-$M$ neighbors for an item in the current graph. Therefore, constructing NSW is much more efficient than constructing an exact k-nearest neighbor graph (knn graph). ip-NSW builds and searches the graph using inner product $s(x,y)=x^{\top}y$ as the similarity function. We omit some details in Algorithm~\ref{alg:nsw query} and Algorithm~\ref{alg:nsw graph construction} for conciseness, for example, ip-NSW actually adopts multiple hierarchical layers of NSW (known as HNSW) to improve performance. Readers may refer to~\cite{HNSW} for more details. 

\begin{algorithm}
	\caption{NSW: Query Processing via Graph Walk~\cite{spreading-out}}
	\label{alg:nsw query}
	\begin{algorithmic}[1]
		\STATE {\bfseries Input:} graph $\mathcal{G}_s$, similarity function $s(x,y)$, query $q$, entry vertex $v_0$, candidate pool size $l$ 
		\STATE Initialize $i=0$, candidate pool $\mathcal{C}=\emptyset$ and $\mathcal{C}$.add($v_0$)
		\WHILE{$i<l$}
		\STATE Set $v_{curr}$ as the first unchecked vertex in $\mathcal{C}$ and set $i$ as its index in $\mathcal{C}$, mark $v_{curr}$ as checked    
		\FOR {every neighbour $v$ of $v_{curr}$ in $\mathcal{G}_s$}
		\STATE  If $v$ is not checked, calculate $s(q, v)$ and $\mathcal{C}$.add($v$) 
		\ENDFOR
		\STATE Sort $\mathcal{C}$ in descending order of $s(q, v)$
		\STATE If $\mathcal{C}$.size()$>$$l$, execute $\mathcal{C}$.resize($l$) by removing items with small $s(q, v)$ 
		\ENDWHILE
		\RETURN the top $k$ vertexes in $\mathcal{C}$  
	\end{algorithmic}
\end{algorithm} 

\begin{algorithm}
	\caption{NSW: Graph Construction~\cite{morozov:graphmips}}
	\label{alg:nsw graph construction}
	\begin{algorithmic}[1]
		\STATE {\bfseries Input:} dataset $\mathcal{X}$, similarity function $s(x,y)$, maximum vertex degree $M$ 
		\STATE Initialize $\mathcal{G}_s=\emptyset$
		\FOR {each $x$ in $\mathcal{X}$ }
		\STATE Use Algorithm~\ref{alg:nsw query} to find $M$ items most similar to $x$ w.r.t.  $s(x,y)$ in the current graph $\mathcal{G}_s$ 
		\STATE Add $x$ to  $G_s$ by connecting it to the $M$ items using directed edges 
		\ENDFOR
		\STATE return $\mathcal{G}_s$ 
	\end{algorithmic}
\end{algorithm}

\subsection{Norm Bias in ip-NSW}\label{sec:graph}

\begin{figure*}[!t]
	\centering
	\includegraphics[width=0.24\textwidth]{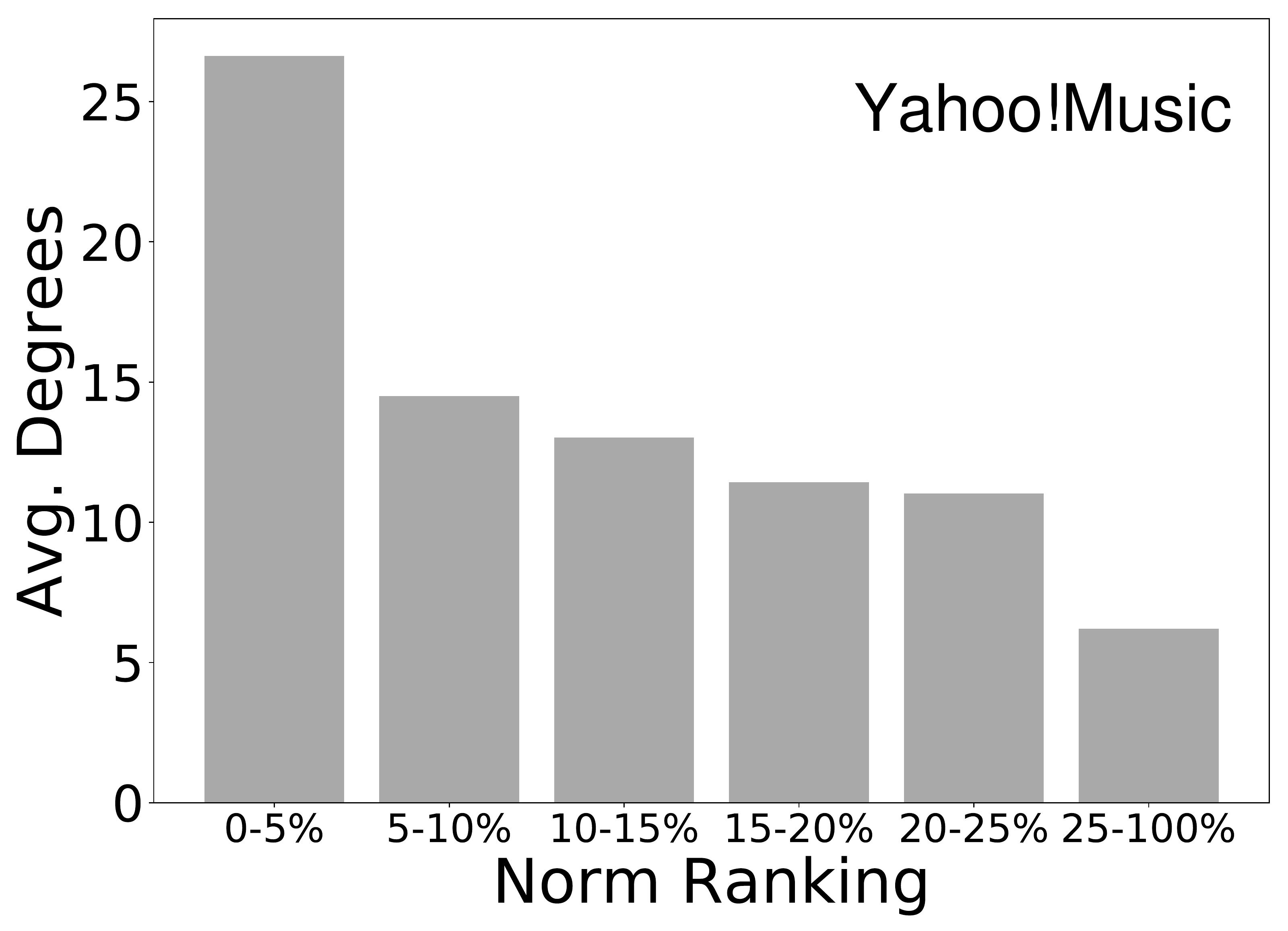}
	\includegraphics[width=0.24\textwidth]{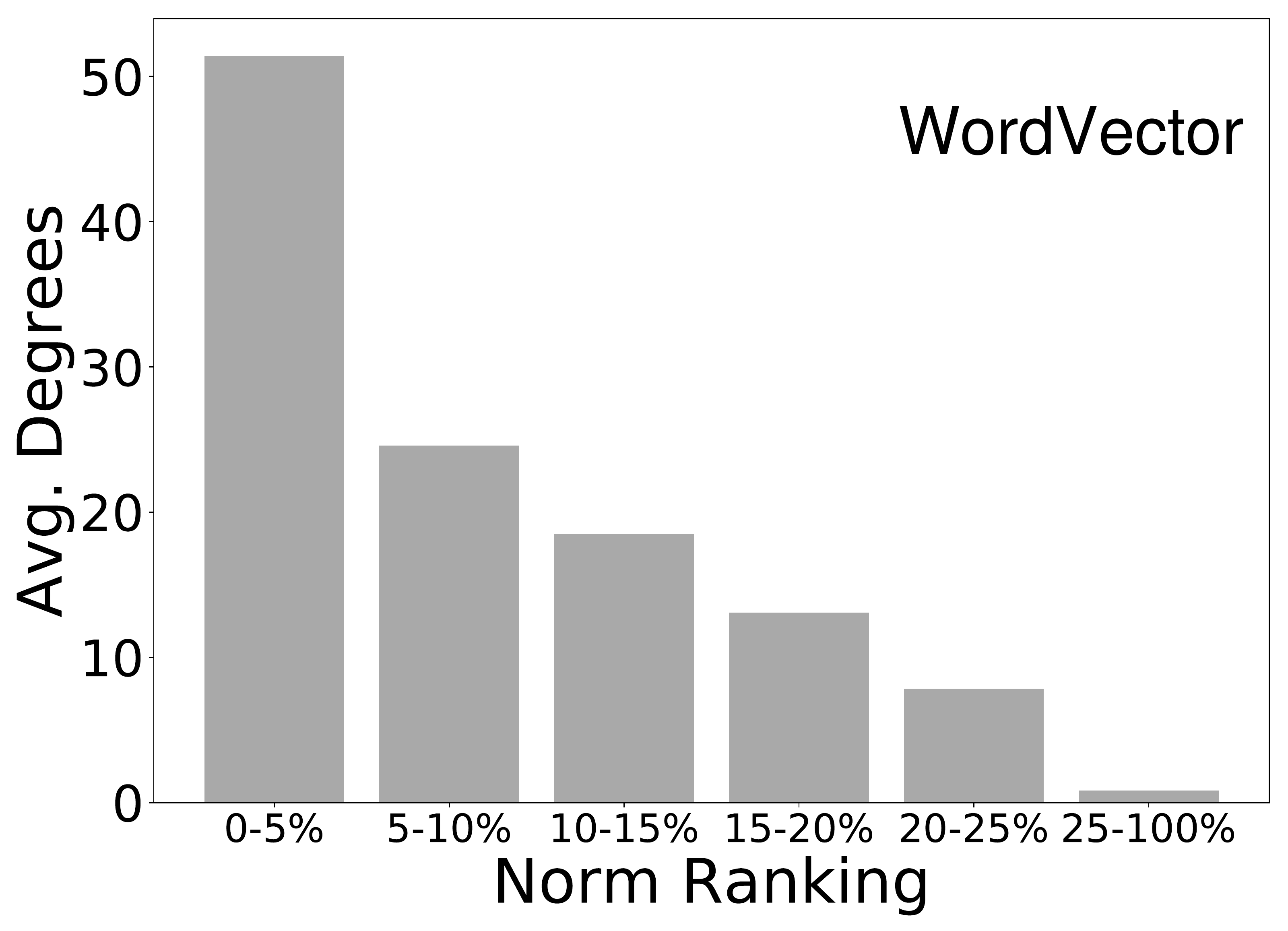}
	\includegraphics[width=0.24\textwidth]{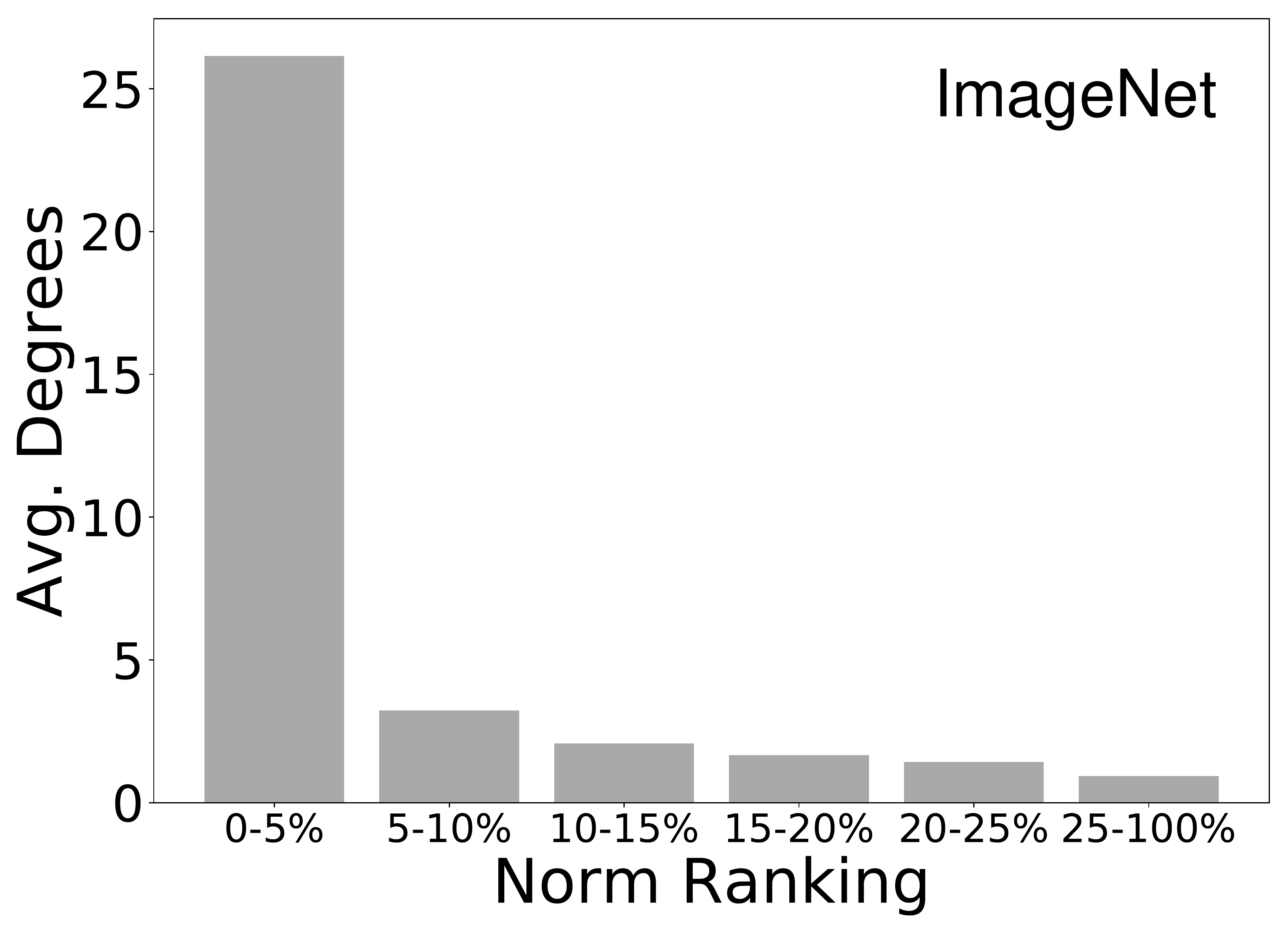}
	\includegraphics[width=0.24\textwidth]{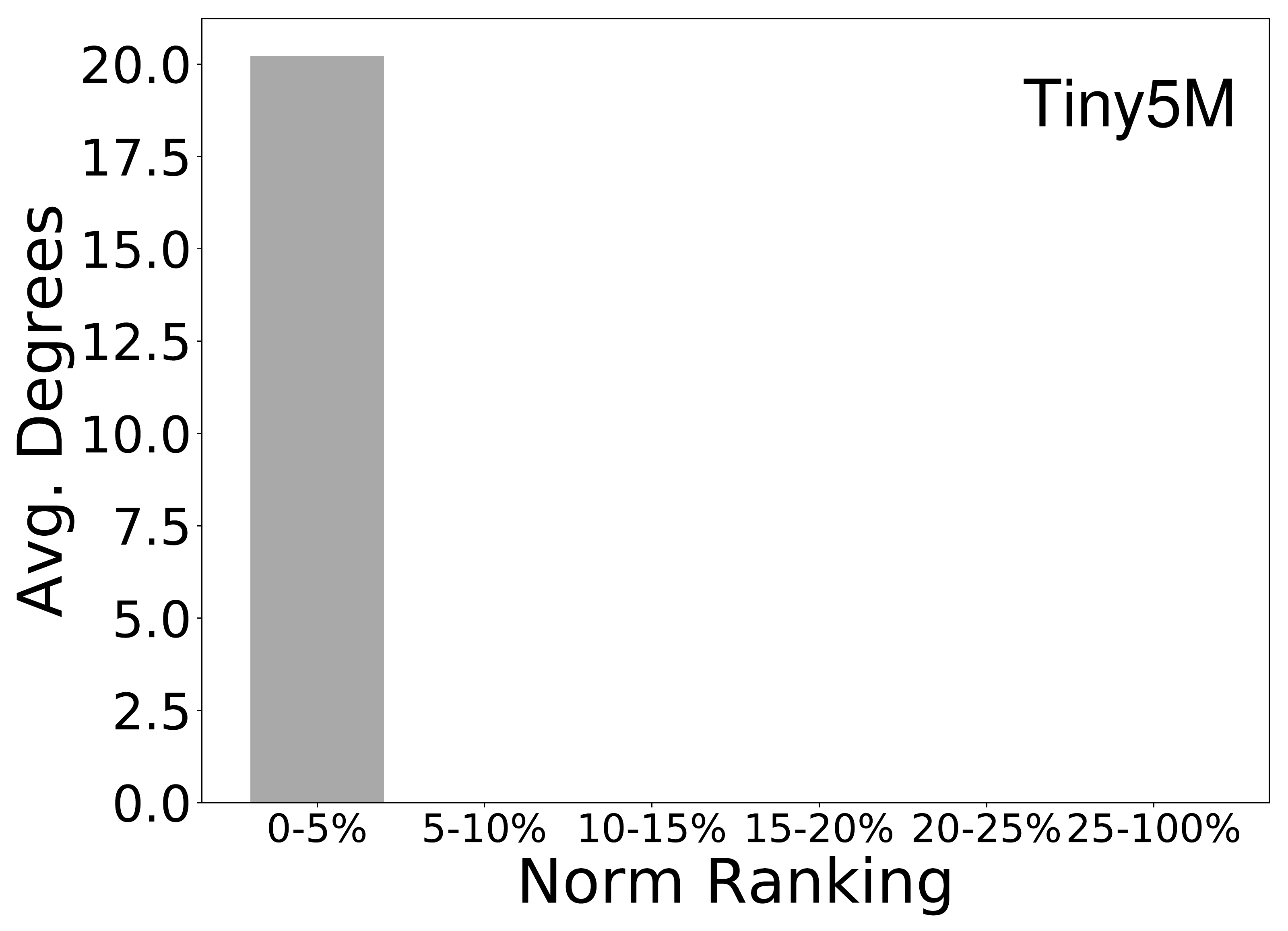}	
	\caption{The average in-degree distribution for items in each norm group}
	\label{fig:in-degree distribution}
\end{figure*}

We built ip-NSW graphs for the four datasets in Table~\ref{tab:datasets} and plot the average in-degree for items in each norm group in Figure~\ref{fig:in-degree distribution}. The results show that the large norm items have much higher in-degrees than the average. To be more specific, the average in-degrees for items ranking top 5\% in norm are 3.2, 8.0, 11.1 and 19.8 times of the dataset average for Yahoo!Music, WordVector, ImageNet and Tiny5M, respectively. This is not surprising as the large norm items are more likely to have large inner product with other items as shown in Section~\ref{sec:norm bias}. The insertion based graph construction procedure of ip-NSW may also contribute to the skewed in-degree distribution. A new item builds its connections by checking the neighbors of existing items and the initially inserted items are likely to connect to the large norm items, which means that graph construction tend to amplify the in-degree skewness. Having large in-degrees means that the large norm items are well-connected in the ip-NSW graph, which makes it more likely for a graph walk to reach them.         

\begin{figure*}[!t]
	\centering
	\includegraphics[width=0.24\textwidth]{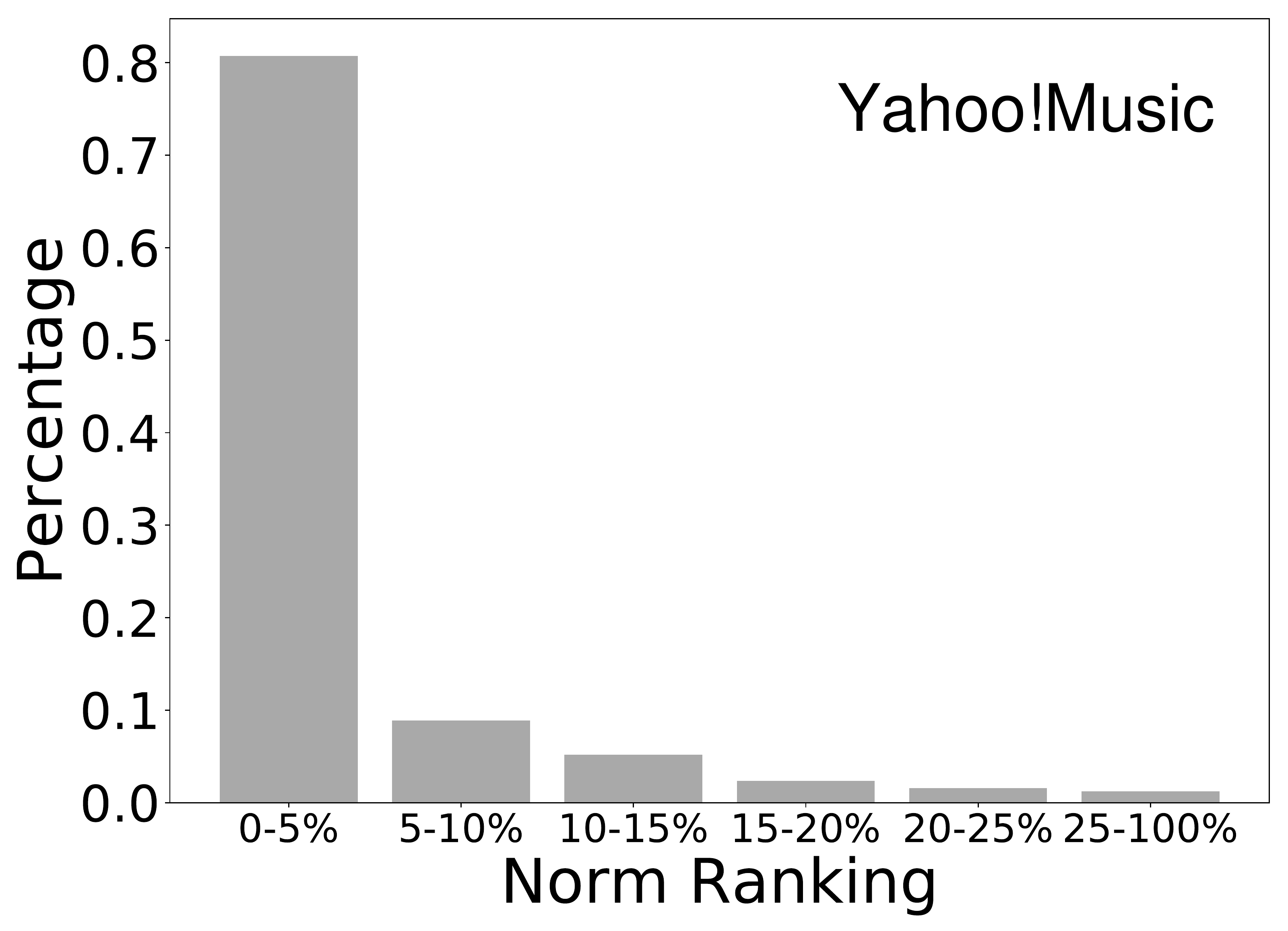}
	\includegraphics[width=0.24\textwidth]{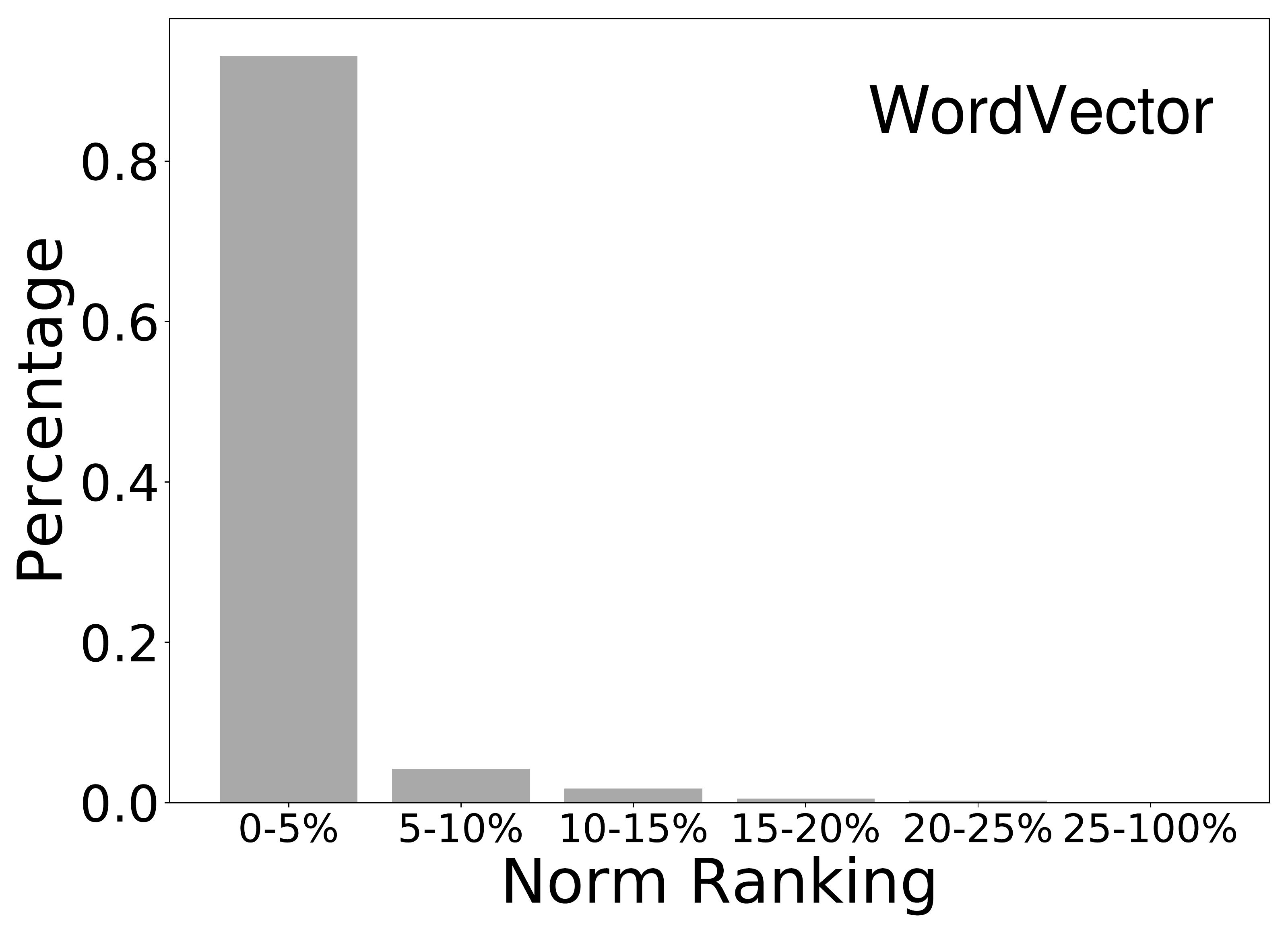}
	\includegraphics[width=0.24\textwidth]{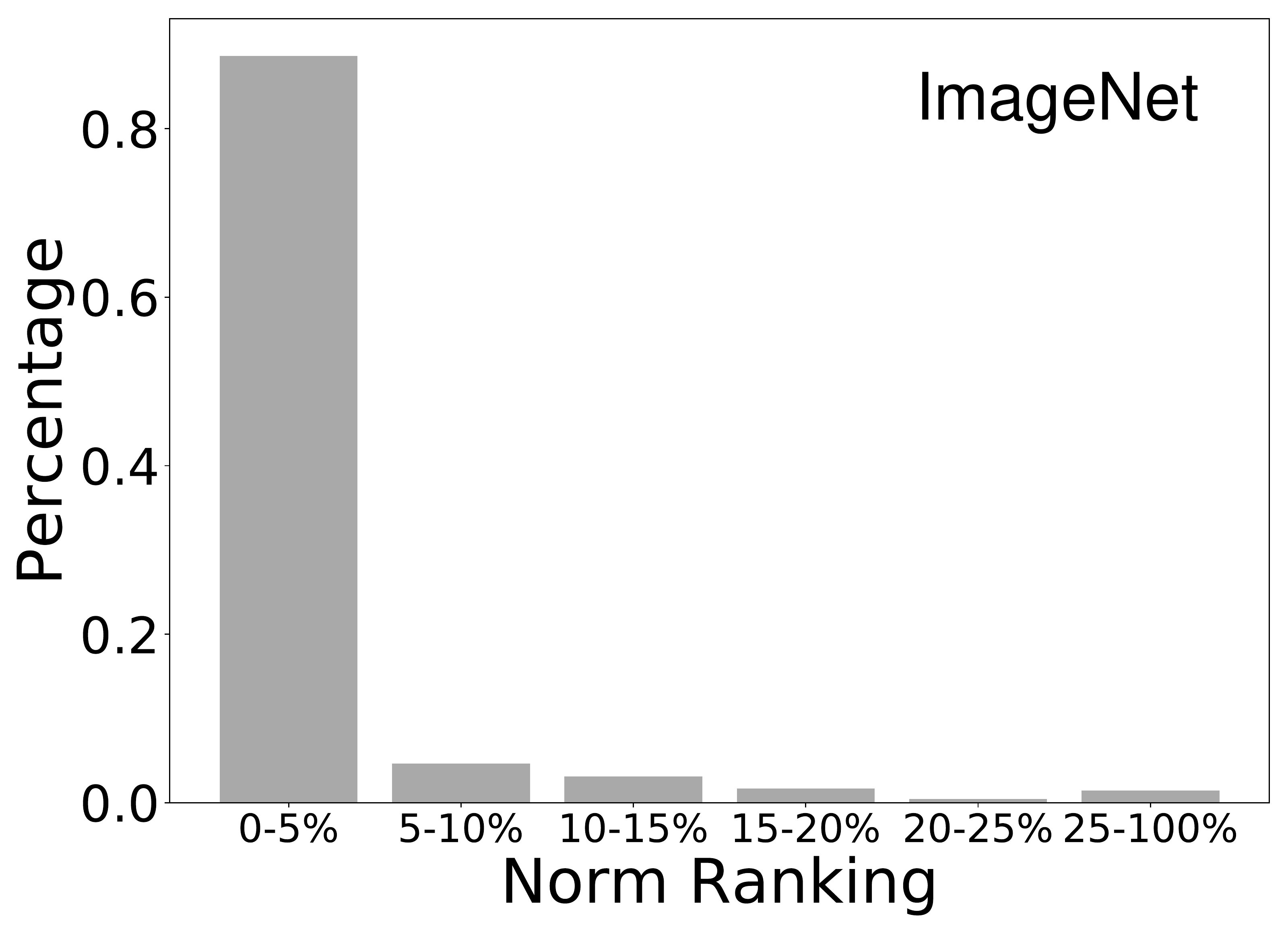}
	\includegraphics[width=0.24\textwidth]{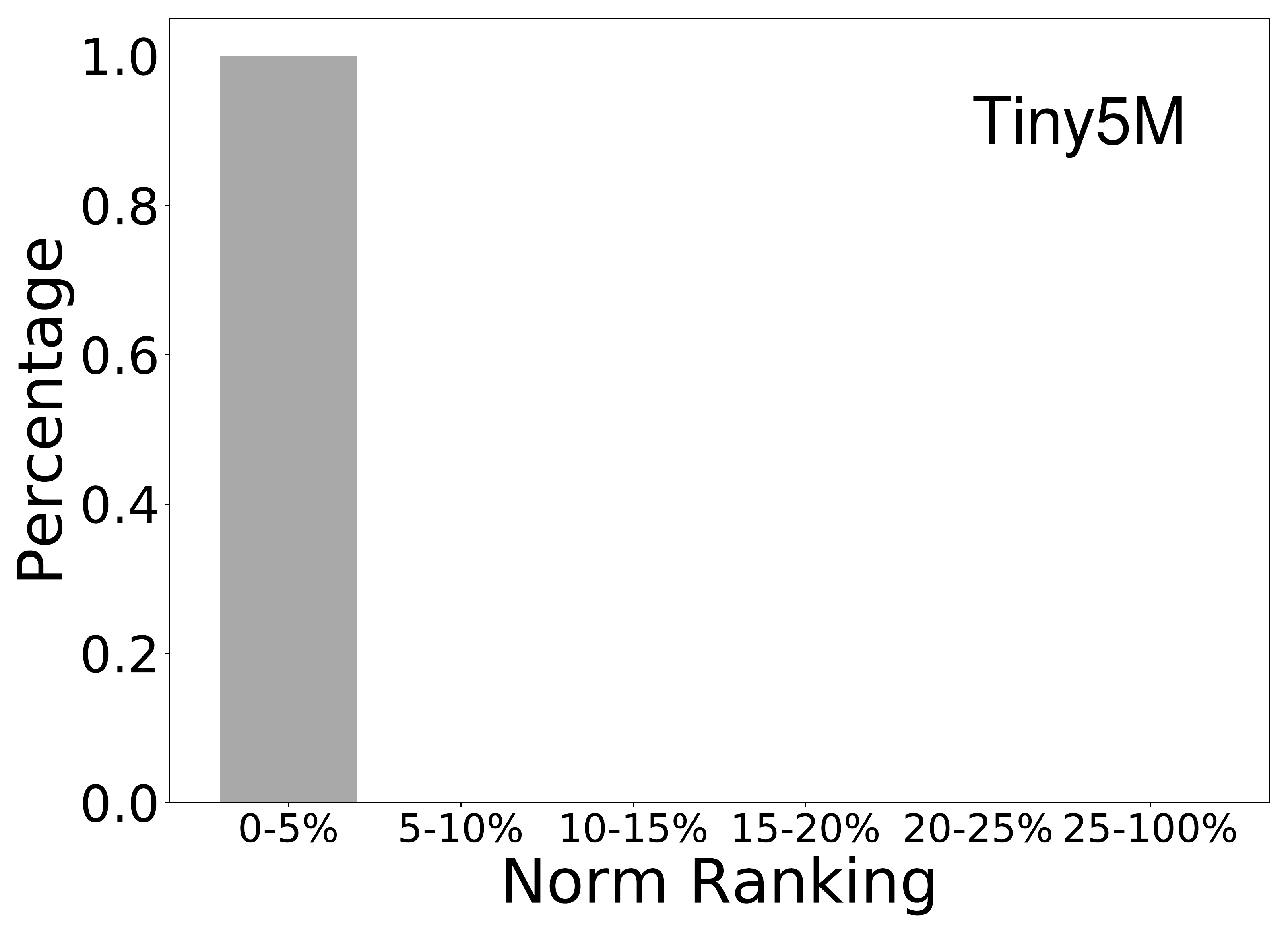}
	\caption{The percentage of inner product computation conducted on items in each norm group}
	\label{fig:computation}
\end{figure*}

To better understand a walk in the ip-NSW graph, we conducted MIPS using ip-NSW for 1,000 randomly selected queries. We recorded the id of the item when inner product was computed, and plot the percentage of inner product computation conducted on items in each norm group in Figure~\ref{fig:computation}. The results show that most of the inner product computation was conducted on the large norm items. For Yahoo!Music, WordVector, ImageNet and Tiny5M, items ranking top 5\% in norm take up 80.7\%, 93.1\%, 88.6\% and 100\% of the inner product computation. Compared with the in-degree distributions in Figure~\ref{fig:in-degree distribution}, the computation distributions are even more biased towards the large norm items. This suggests that a walk in the ip-NSW graph reaches the large norm items very quickly and keeps moving among these items. With these results, we can conclude that ip-NSW is also biased towards the large norm items, in terms of both connectivity and computation. The norm bias of ip-NSW allows it to effectively avoid unnecessary computation on small norm items that are unlikely to be the result of MIPS. Therefore, ip-NSW has excellent performance mainly because it matches the strong norm bias of the MIPS problem. 

\section{The ip-NSW+ Algorithm}\label{sec:method}

In this section, we present the ip-NSW+ algorithm, which is motivated by an analysis indicating that the norm bias of ip-NSW can lead to inefficient MIPS.

\subsection{Motivation}\label{sec:motivation}

\begin{figure}[!t]	
	\centering 
	\begin{minipage}[b]{0.23\textwidth}
		\centering
		\includegraphics[width=\textwidth]{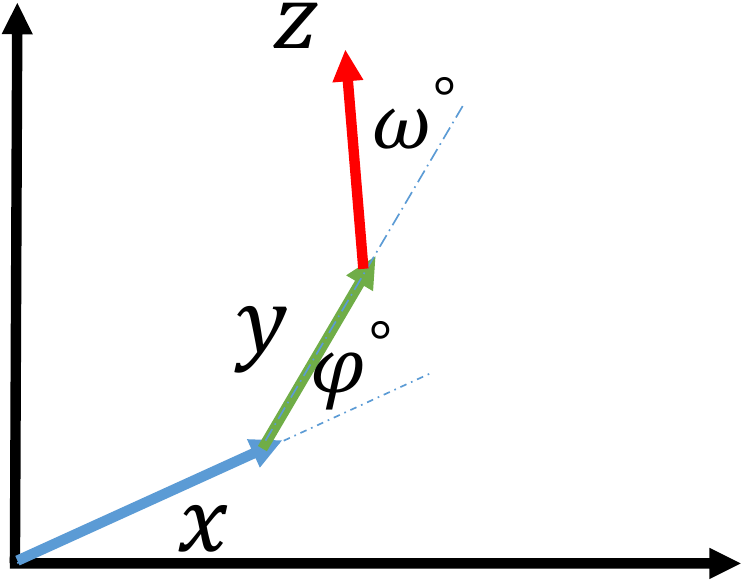}
		\subcaption{MIPS neighbour}\label{fig:MIPS neigbour}
	\end{minipage}
	\begin{minipage}[b]{0.23\textwidth}
		\centering
		\includegraphics[width=\textwidth]{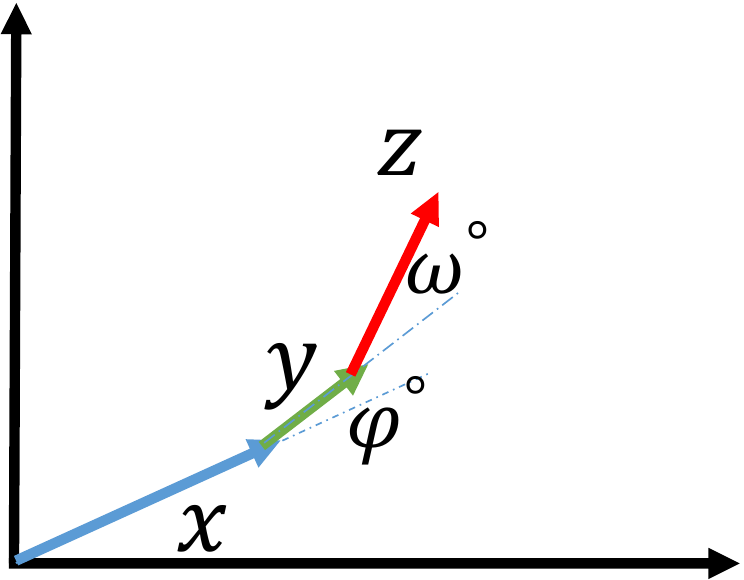}
		\subcaption{Angular neighbour}\label{fig:Angular neigbour}
	\end{minipage}
	\caption{Example of MIPS neighbor and angular neighbor}
	\label{fig:imagenet change}
\end{figure}

We have shown in Section~\ref{sec:ip-NSW} that ip-NSW has a strong norm bias, which helps to avoid computation on small norm items. However, this norm bias can result in inefficient MIPS and we illustrate this point with an example in Figure~\ref{fig:MIPS neigbour}, in which $y$ is an MIPS neighbor of $x$ and $z$ is an MIPS neighbor of $y$. As $y$ and $z$ are the MIPS neighbors of some item, they usually have large norm due to the norm bias of the MIPS problems but the angles ($\varphi$ and $\omega$) are not necessarily small, especially when the norm of $y$ and $z$ are very large. Suppose that $x$ is the query and the graph walk is now at $y$, in the next step, the graph walk will compute $x^{\top}z$ but $z$ may not have a good inner product with $x$ due to the large angle (i.e., $\varphi + \omega$) between them. This example shows that ip-NSW may spend computation on many large norm items that do not have a good inner product with the query because the large norm items are well connected in the ip-NSW graph. 

The problem of ip-NSW is caused by the rule it adopts --- the MIPS neighbor of an MIPS neighbor is also likely to be an MIPS neighbor, which is not necessarily true. To improve ip-NSW, we need a new rule that satisfies two requirements. First, it should match the norm bias of the MIPS problems and avoid computation on small norm items, which ip-NSW does well. Second, it should also avoid computation on large norm items that do not have a good inner product with the query, which is the main problem of ip-NSW.       

We propose an alternative rule --- the MIPS neighbor of an angular neighbor is likely to be an MIPS neighbor, which satisfies the two requirements. We define the angular similarity between two vector $x$ and $y$ as $s_a(x, y)\!=\!\frac{x^{\top}y}{\Vert x\Vert \Vert y\Vert}$ and say that $y$ is an angular neighbor of $x$ if $s_a(x, y)$ is large. Specifically, this rule says that for a query $x$ and its angular neighbor $y$ in a dataset, if $z$ is an MIPS neighbor of $y$ in the dataset, then $x^{\top}z$ is likely to be large. We provide an illustration of this rule in Figure~\ref{fig:Angular neigbour}. In the figure, $z$ is an MIPS neighbor of $y$, thus $z$ usually has large norm, meeting the first requirement. The angle $\varphi + \omega$ is usually not too large as $y$ is an angular neighbor of $x$ and $\varphi$ is small, and thus $x^{\top}z$ is likely to be large, meeting the second requirement. Theorem~\ref{theorem:ip} formally establishes that an MIPS neighbor of an angular neighbor is a good MIPS neighbor with an assumption about $z$.       

\begin{theorem}\label{theorem:ip}
	For two vectors $x$ and $y$ in $\mathbb{R}^d$ having an angular similarity $\beta\!=\!\frac{x^{\top}y}{\Vert x \Vert \Vert y \Vert}$, a third vector $z\in\mathbb{R}^d$ and the entries of $z$ are independent and $z_i\!\sim\!\mathcal{N}(0,1)$ for $i=1,2,\cdots,d$, given $y^{\top}z=\gamma$, we have $x^{\top}z  \mid y^{\top}z=\gamma \sim\mathcal{N}(\frac{\gamma \beta \Vert x \Vert}{\Vert y \Vert},\Vert x \Vert^2(1-\beta^2))$. 
\end{theorem}

The proof can be found in the supplementary material. If $x$ is a query and $y$ is the angular neighbor of $x$ in the dataset, which means that $\beta\!=\!\frac{x^{\top}y}{\Vert x \Vert \Vert y \Vert}$ is large. If $y$ and $z$ are both in the dataset and $z$ is an MIPS neighbor of $y$, we have $y^{\top}z=\rho \Vert y \Vert \Vert z \Vert$, in which $\rho$ is large. Given these conditions and using Theorem~\ref{theorem:ip}, we have $x^{\top}z  \mid y^{\top}z=\rho \Vert y \Vert \Vert z \Vert \sim\mathcal{N}(\rho \beta \Vert x \Vert \Vert z \Vert,\Vert x \Vert^2(1-\beta^2))$, which means the inner product between $x$ and $z$ follows a Gaussian distribution. The mean of the distribution ($\rho \beta \Vert x \Vert \Vert z \Vert$) is large as both $\rho$ and $\beta$ are large. Moreover, the variance of the distribution ($\Vert x \Vert^2(1-\beta^2)$) is small as $\beta$ is large. Therefore, there is a good chance that $x^{\top}z$ is a large.       

Theorem~\ref{theorem:ip} is also supported by empirical results from the following experiment. We conducted search for 1,000 randomly selected queries on Yahoo!Music and ImageNet. For each query, we find its ground truth top-10 angular neighbors in the dataset and for each of these angular neighbors, we find its ground truth top-10 MIPS neighbors in the dataset. This procedure gives us a result set containing 100 candidates (with possible duplication) for each query, which can be used to calculate the recall for top-10 MIPS. The average recalls were 82.67\% and 97.22\% for Yahoo!Music and ImageNet, respectively, which suggests that aggregating the MIPS neighbors of the angular neighbors can obtain a good recall for MIPS. In contrast, aggregating the MIPS neighbors of the ground-truth top-10 MIPS neighbors of a query only provide a recall of 67.21\% for ImageNet.

\subsection{ip-NSW+}

Based on the new rule presented in Section~\ref{sec:motivation}, we present the query processing procedure of ip-NSW+ in Algorithm~\ref{alg:ip-NSW+ graph construction}.

\begin{algorithm}
	\caption{ip-NSW+: Query Processing via Graph Walk}
	\label{alg:ip-NSW+ graph construction}
	\begin{algorithmic}[1]
		\STATE {\bfseries Input:} an angular NSW graph $\mathcal{A}_s$, an inner product NSW graph $\mathcal{G}_s$, query $q$, starting vertex $v_0$ in $\mathcal{A}_s$
		\STATE Conduct search on $\mathcal{A}_s$ using Algorithm~\ref{alg:nsw query} to find the top-$k'$ angular neighbors of $q$
		\FOR {each item $v$ in the top-$k'$ angular neighbors}
		\FOR {each edge $(v, u)$ in the ip-NSW graph $\mathcal{G}_s$}
		\STATE $\mathcal{C}$.add($u$) 
		\ENDFOR
		\ENDFOR
		\STATE Conduct search on $\mathcal{G}_s$ with $\mathcal{C}$ using Algorithm~\ref{alg:nsw query} to find the top-$k$ inner product neighbors of $q$     
	\end{algorithmic}
\end{algorithm}

To find the angular neighbors of the query, ip-NSW+ searches an angular NSW graph $\mathcal{A}_s$ because NSW provides excellent performance on many similarity search problems. Instead of finding the exact inner product neighbors of the angular search results, ip-NSW+ uses their neighbors in the inner product graph $\mathcal{G}_s$ as an approximation. After the initialization (line 2-5 in Algorithm~\ref{alg:ip-NSW+ graph construction}), the candidate queue $\mathcal{C}$ already contains a good portion of the MIPS result for the query and the time spent to find them by graph walk on ip-NSW can be saved. To further refine the result in $\mathcal{C}$, a standard graph walk on the inner product graph $\mathcal{G}_s$ is conducted in line 6 of Algorithm~\ref{alg:ip-NSW+ graph construction}.

For index construction, ip-NSW+ builds $\mathcal{A}_s$ and $\mathcal{G}_s$ simultaneously and the items are inserted sequentially (in a random order) into the two graphs. For an item $v$, it is first inserted into $\mathcal{A}_s$ with Algorithm~\ref{alg:nsw graph construction} using angular similarity as the similarity function. Then, $v$ is inserted into $\mathcal{G}_s$ and the neighbors of $v$ in $\mathcal{G}_s$ are found using ip-NSW+ (Algorithm~\ref{alg:ip-NSW+ graph construction}) instead of ip-NSW (Algorithm~\ref{alg:nsw query}). Empirically, we found that this provides more accurate inner product neighbors for the items and hence better search performance. One subtlety of ip-NSW+ is controlling the time spent on angular neighbor search (ANS). Spending too much time for ANS means only a short time is left for result refinement by a graph walk on the inner product graph $\mathcal{G}_s$, which harms performance. As the time consumption of a graph walk in NSW is controlled by the maximum degree $M$ (the complexity of each step) and the candidate pool size $l$ (how many steps will be taken), we use smaller $M$ and $l$ for the angular graph $\mathcal{A}_s$ than for the inner product graph $\mathcal{G}_s$. We show in Section~\ref{sec:exp} that ip-NSW+ using fixed $M$ and $l$ without dataset-specific tuning already performs significantly better than ip-NSW.

The index construction complexity of ip-NSW+ is approximately twice of ip-NSW as ip-NSW+ constructs two proximity graphs. The index size of ip-NSW+ is less than twice of ip-NSW because we use small $M$ for the angular graph $\mathcal{A}_s$. These additional complexities will not be a big problem because the insertion-based graph construction of NSW is efficient and the memory of a single machine is sufficient for most datasets. However, ip-NSW+ provides significantly better recall-time performance than ip-NSW (see Section~\ref{sec:exp}), which benefits many applications. Existing proximity-graph-based algorithms use a single proximity graph and the same similarity function is used for both index construction and query processing. In contrast, ip-NSW+ uses two proximity graphs constructed from different similarity functions jointly, which is a new paradigm for proximity-graph-based similarity search and may inspire future research.              

\section{Experimental Results}\label{sec:exp}

\begin{figure*}[!t]
	\centering
	\includegraphics[width=0.24\textwidth]{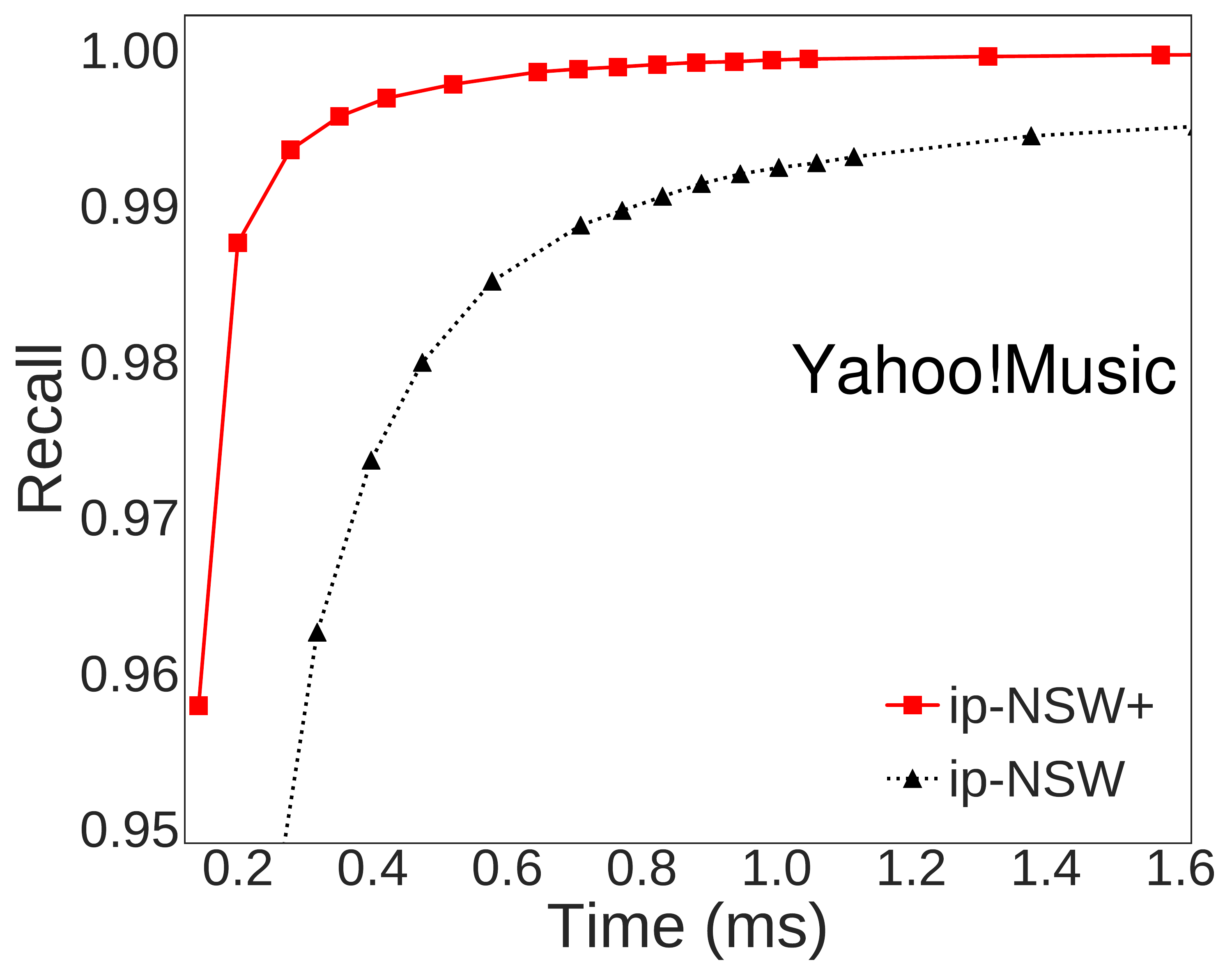}
	\includegraphics[width=0.24\textwidth]{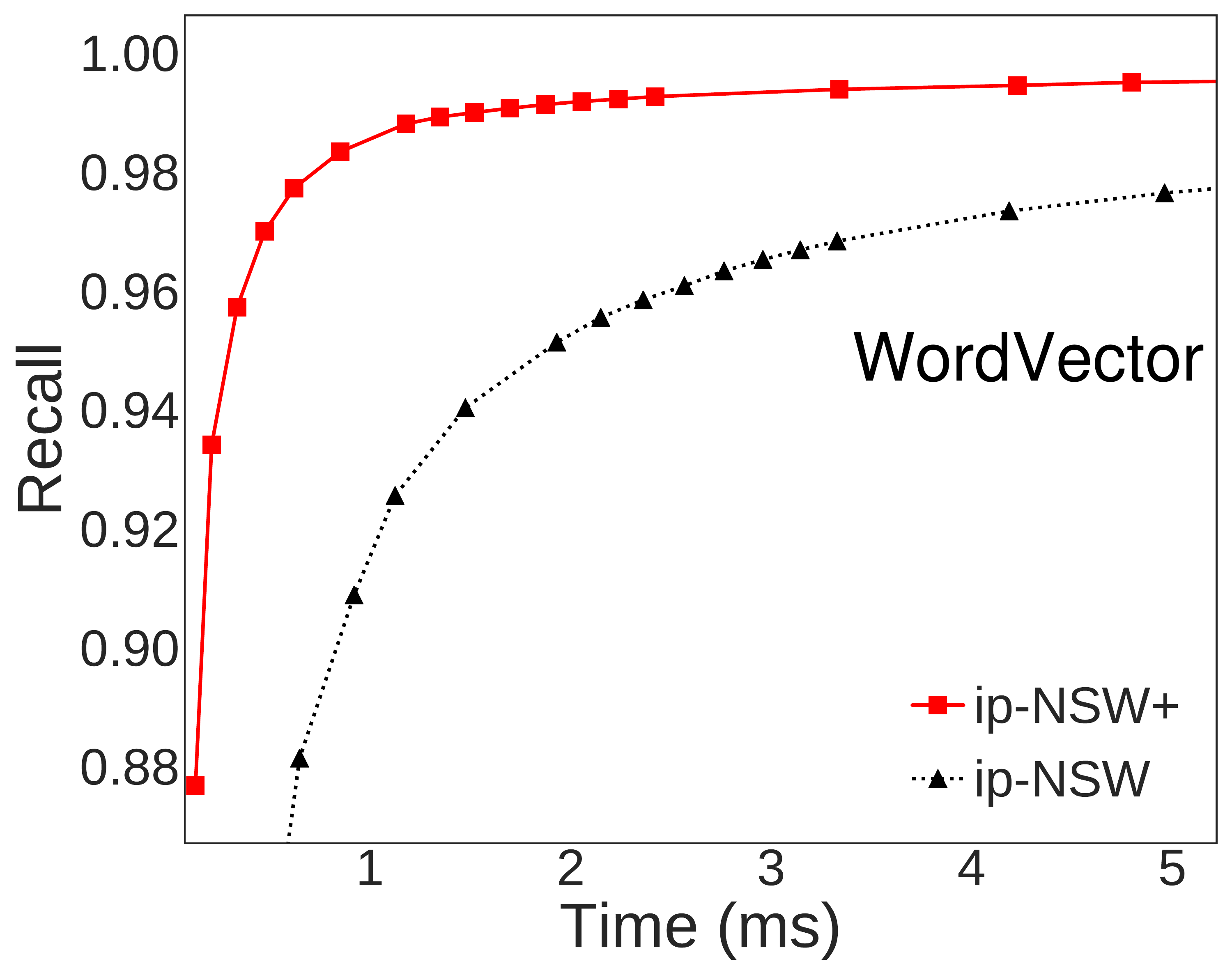}
	\includegraphics[width=0.24\textwidth]{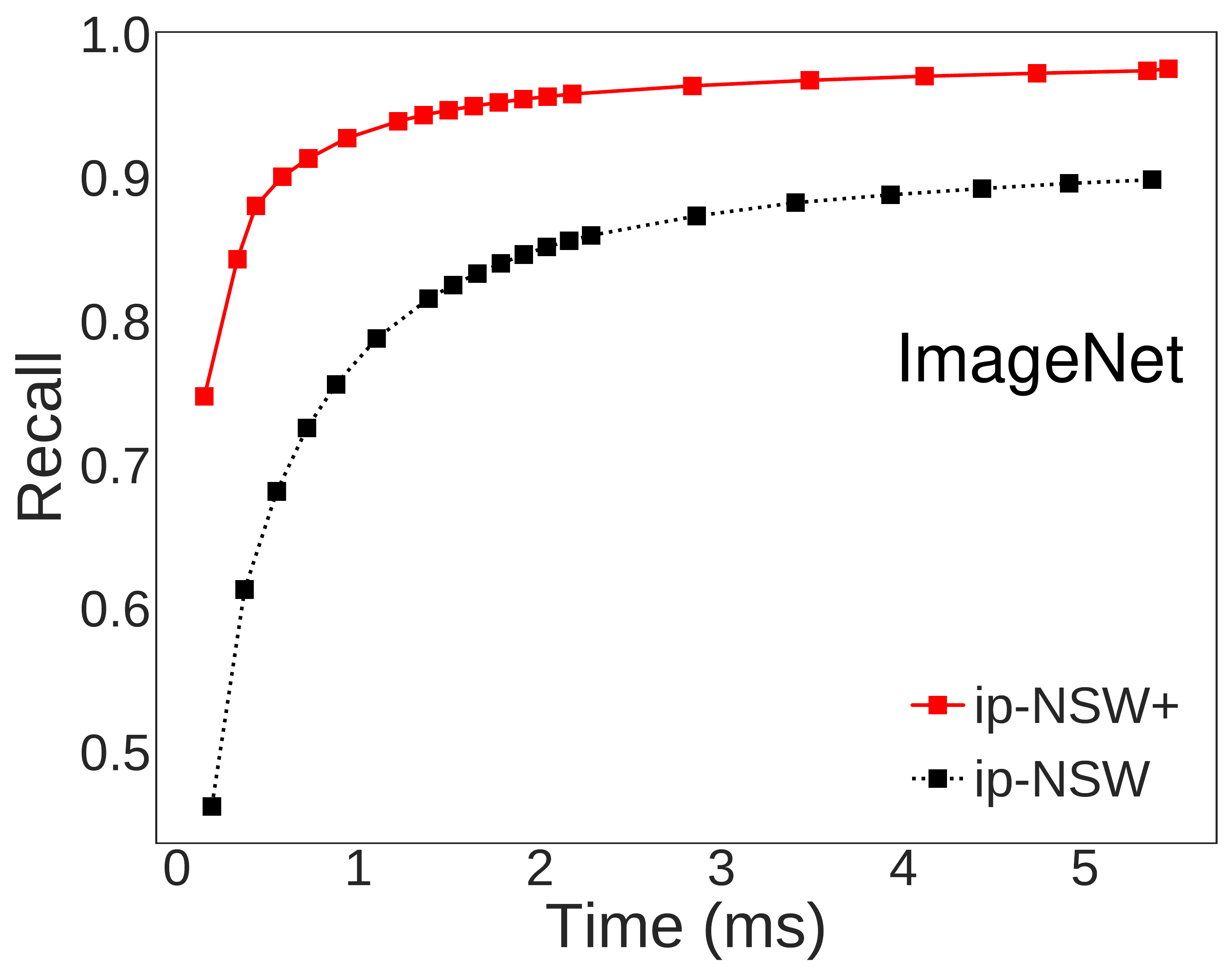}
	\includegraphics[width=0.24\textwidth]{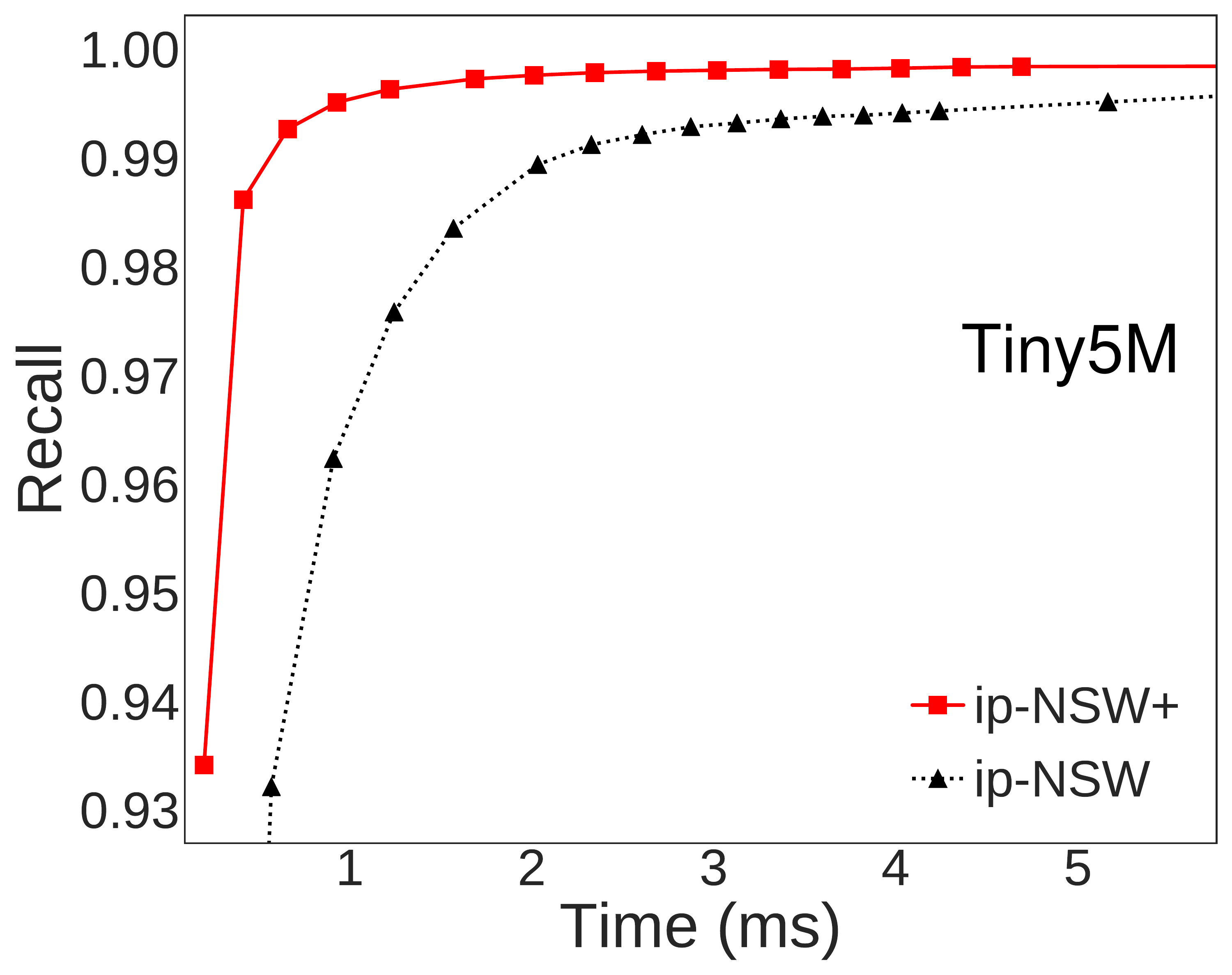}	
	\caption{Recall-time performance comparison between ip-NSW and ip-NSW+ }
	\label{fig:main result}
\end{figure*}

\begin{figure*}[!h]	
	\centering 
	\begin{minipage}[b]{0.24\textwidth}
		\centering
		\includegraphics[width=\textwidth]{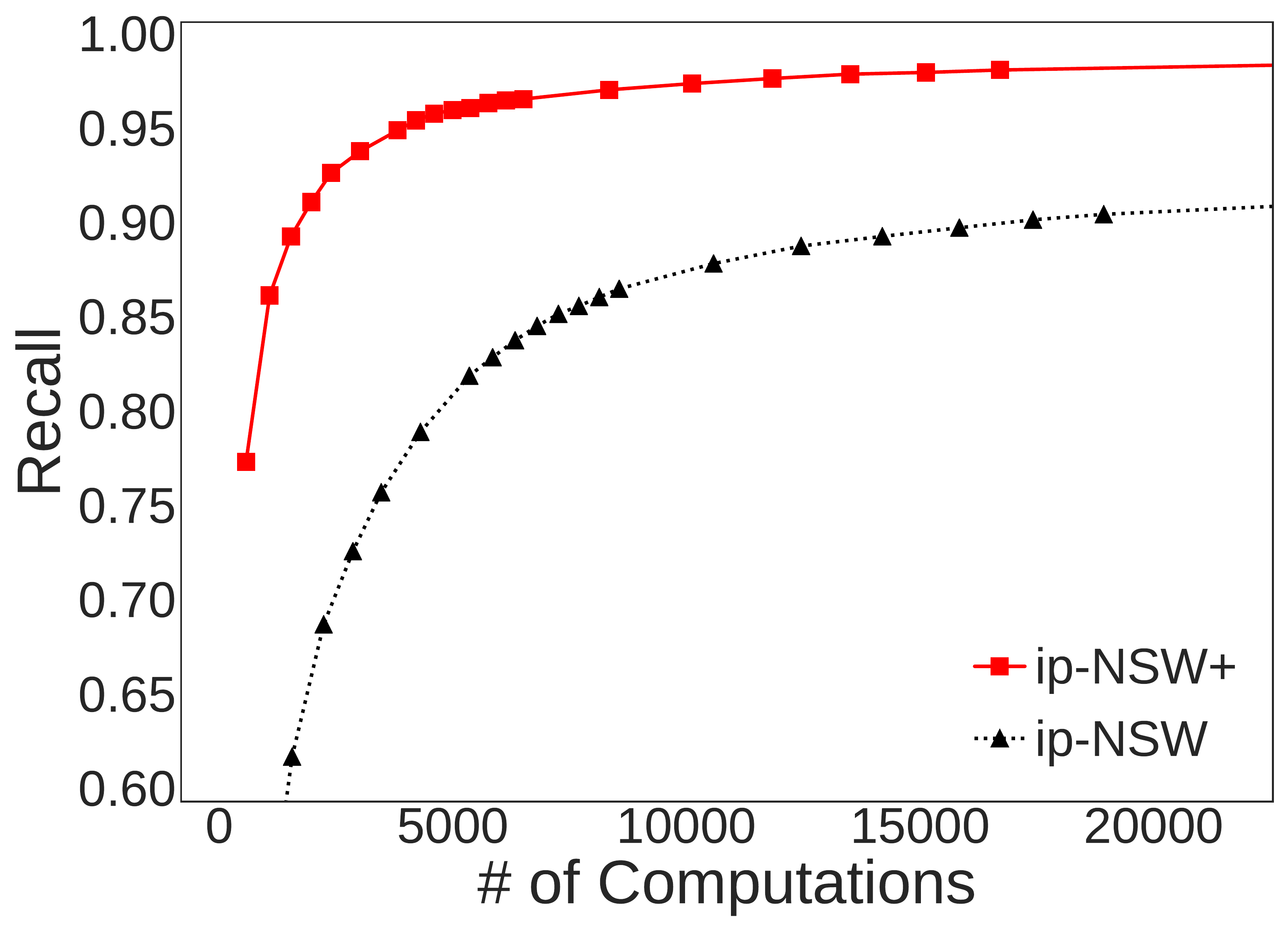}
		\subcaption{\# computation vs. recall}\label{fig:yahoomusic-u}
	\end{minipage}
	\begin{minipage}[b]{0.24\textwidth}
		\centering
		\includegraphics[width=\textwidth]{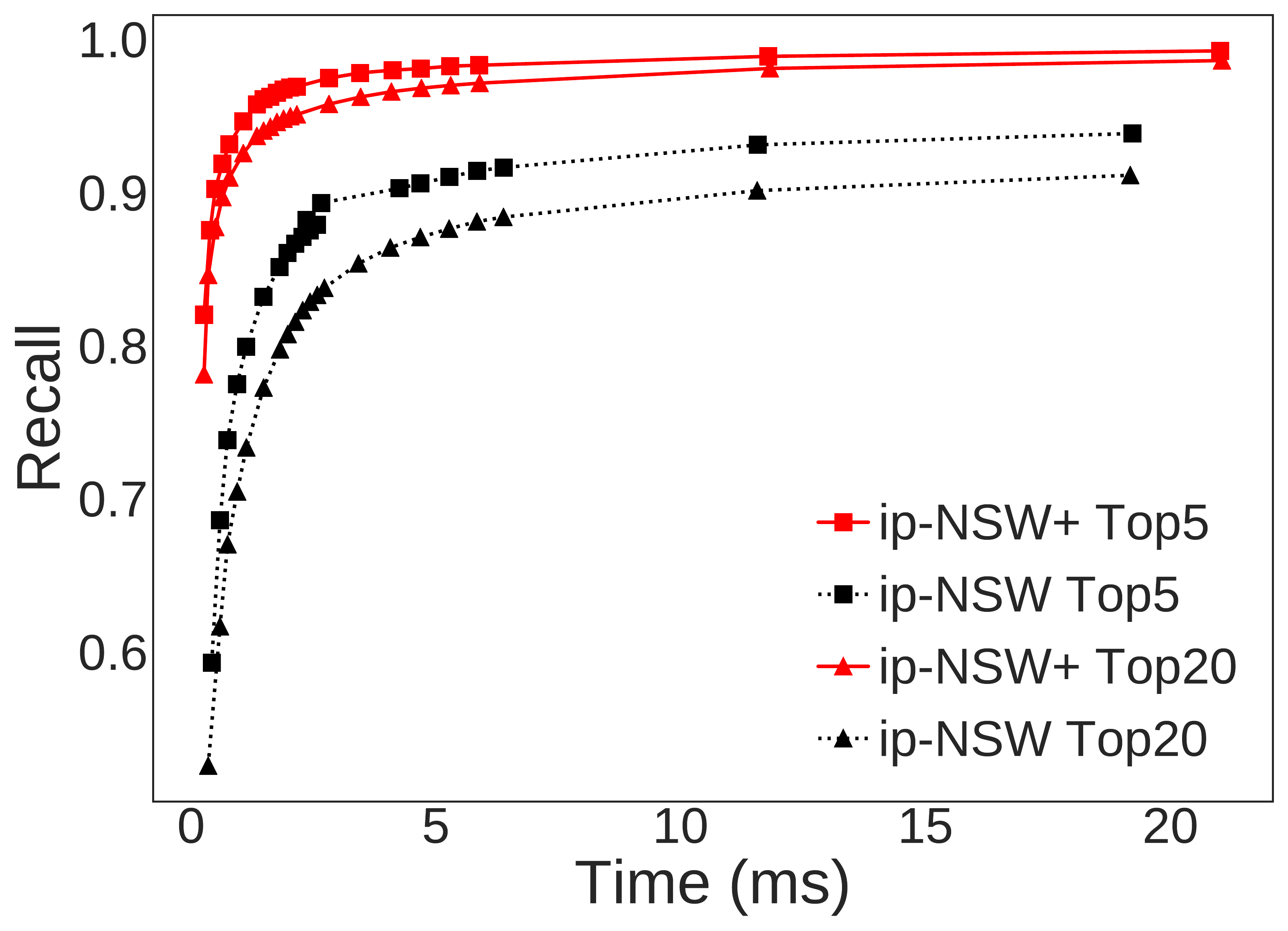}
		\subcaption{Different values of $k$}\label{fig:imagenet-u}
	\end{minipage}
	\begin{minipage}[b]{0.24\textwidth}
		\centering
		\includegraphics[width=\textwidth]{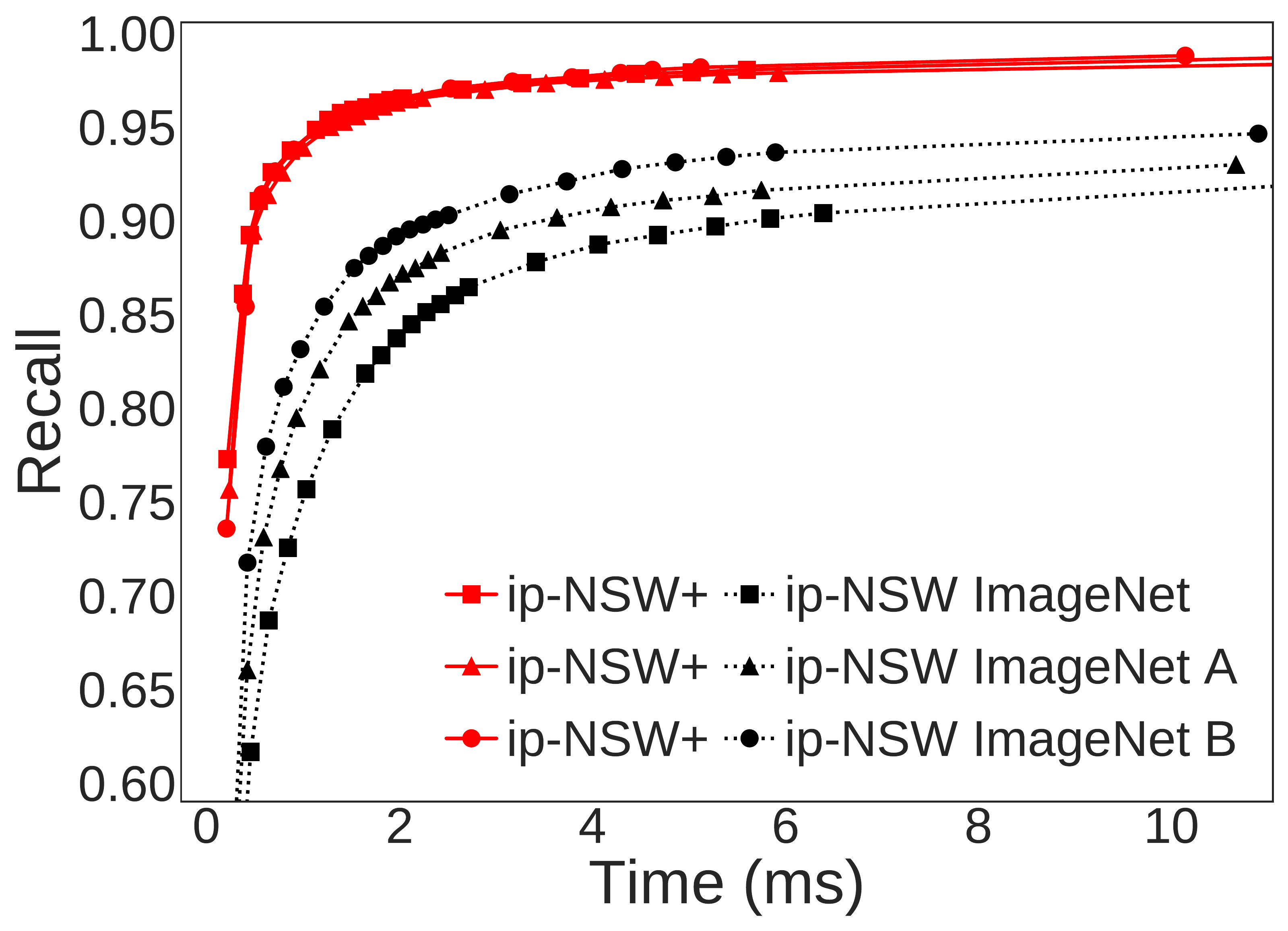}
		\subcaption{ImageNet Variants}\label{fig:imagenet-scaling}
	\end{minipage}
	\caption{Additional experimental results on the ImageNet dataset (best viewed in color)}
	\label{fig:more}
\end{figure*}

\textbf{Datasets and settings.} We used the four datasets listed in Table~\ref{tab:datasets}. Yahoo!Music is obtained by conducting ALS-based matrix factorization~\cite{yun:als} on the user-item ratings in the Yahoo!Music dataset. We used the item embeddings as dataset items and the user embeddings as queries. WordVector is sampled from the word2vec embeddings released in~\cite{word2vec}, and ImageNet contains the visual descriptors of the ImageNet images~\cite{deng2009imagenet}. Tiny5M is sampled from the Tiny80M dataset and contains visual descriptors of the Tiny images. Unless otherwise stated, we test the performance of top-10 MIPS and use recall as the performance metric. For top-$k$ MIPS, an algorithm only returns the $k$ best items it finds. Denote the results an algorithm returns for a query as $\mathbb{S}'$ and the ground truth top-$k$ MIPS of the query as $\mathbb{S}$, recall is defined as $r=|\mathbb{S}'\cap \mathbb{S}|/|\mathbb{S}|$. We report the average recall of 10,000 randomly selected queries. We used $M=10$ and $l=10$ for the angular graph $\mathcal{A}_s$ in ip-NSW+ in all experiments and the parameter configurations of $\mathcal{G}_s$ in ip-NSW+ is the same as the inner product graph in ip-NSW. The experiments were conducted on a machine with Intel Xeon E5-2620 CPU and 48 GB memory in a single-thread mode. For ip-NSW+, the reported time includes searching both the angular graph $\mathcal{A}_s$ and the inner product graph $\mathcal{G}_s$.  We implemented ip-NSW+ by modifying the code of ip-NSW and did not introduce extra optimizations to make ip-NSW+ run faster~\footnote{See https://github.com/jerry-liujie/ip-nsw/tree/GraphMIPS for all experiment code and data.}. 


\textbf{Direct comparison.} We report the recall-time performance of ip-NSW and ip-NSW+ in Figure~\ref{fig:main result}. We also tested Simple-LSH~\cite{neyshabur:simple-lsh}, the state-of-the-art LSH-based method for MIPS. We used the implementation provided in~\cite{yu2017greedy} and tuned the parameters following~\cite{morozov:graphmips}. However, the performance of Simple-LSH is significantly poorer and plotting its recall-time curve in Figure~\ref{fig:main result} will make the figure hard to read, and thus we report its curve in the supplementary material. As an example, Simple-LSH takes 598ms to reach a recall of 0.732 for WordVector and 1035ms to reach a recall of 0.735 for ImageNet. This is actually worse than the exact MIPS method, FEXIPRO~\cite{li:fexipro}, which uses multiple pruning rules to speed up linear scan, and takes 20.9ms, 196.3ms and 179.5ms on average for each query on Yahoo!Music, WordVector and ImageNet, respectively~\footnote{We did not provide the performance for FEXIPRO on Tiny5M as it goes out of memory.}. FEXIPRO, however, is at least an order of magnitude slower than ip-NSW and ip-NSW+, as shown in Figure~\ref{fig:main result}, which confirms the results in~\cite{morozov:graphmips} that ip-NSW outperforms existing algorithms. Importantly, ip-NSW+ is able to further make significant improvements over ip-NSW. For example, ip-NSW+ reaches a recall of 0.9 at a speed that is 11 times faster than ip-NSW (0.5 ms vs 5.5 ms) on the ImageNet dataset. Even on the Tiny5M dataset, which has the strongest norm bias and items ranking top 5\% in norm occupy 100\% of top-10 MIPS result, ip-NSW+ still outperforms ip-NSW.

\textbf{More experiments.} We conducted this set of experiments on the ImageNet dataset to further examine the performance of ip-NSW+. In Figure~\ref{fig:yahoomusic-u}, we compare the recall of ip-NSW and ip-NSW+ with respect to the number of similarity function evaluations since similarity function evaluation is usually the most time-consuming part of an algorithm. We count as one similarity function evaluation when ip-NSW computes inner product with one item and ip-NSW+ computes angular similarity or inner product with one item. The results show that ip-NSW+ spends much less computation than ip-NSW for the same recall, suggesting the performance gain of ip-NSW+ indeed comes from a better algorithm design. In Figure~\ref{fig:imagenet-u}, we compare ip-NSW and ip-NSW+ for top-5 MIPS and top-20 MIPS, which shows that ip-NSW+ consistently outperforms ip-NSW for different $k$. 

One surprising phenomenon is that ip-NSW+ provides more robust performance than ip-NSW under different transformations of the norm distribution. We created two variants of the ImageNet dataset, i.e., ImageNet-A and ImageNet-B, by scaling the items without changing their directions. ImageNet-A and ImageNet-B add 0.18 and 0.36 to the Euclidean norm of each item, respectively. The norm distributions of the transformed datasets can be found in the supplementary material. We define the tailing factor (TF) of a dataset as the ratio between the 95\% percentile of the norm distribution and the median norm and say that the norm distribution is more skewed when the TF is large. The TFs of ImageNet, ImageNet-A and ImageNet-B are 2.05, 1.55 and 1.37, respectively. We report the performance of ip-NSW and ip-NSW+ on the three datasets in Figure~\ref{fig:imagenet-scaling}. The results show that ip-NSW+ has almost identical performance on the three ImageNet variants and consistently outperforms ip-NSW. In contrast, the performance of ip-NSW varies a lot, the best performance is achieved on ImageNet-B (with the smallest TF) while the worst performance is observed on ImageNet (with the largest TF). We tried more datasets and another method to scale the items in the supplementary material and the results show that ip-NSW+ consistently provides more robust performance than ip-NSW. Moreover, ip-NSW usually performs better when the TF is small.  

\begin{figure}[!t]	
	\centering 
	\begin{minipage}[b]{0.23\textwidth}
		\includegraphics[width=\textwidth]{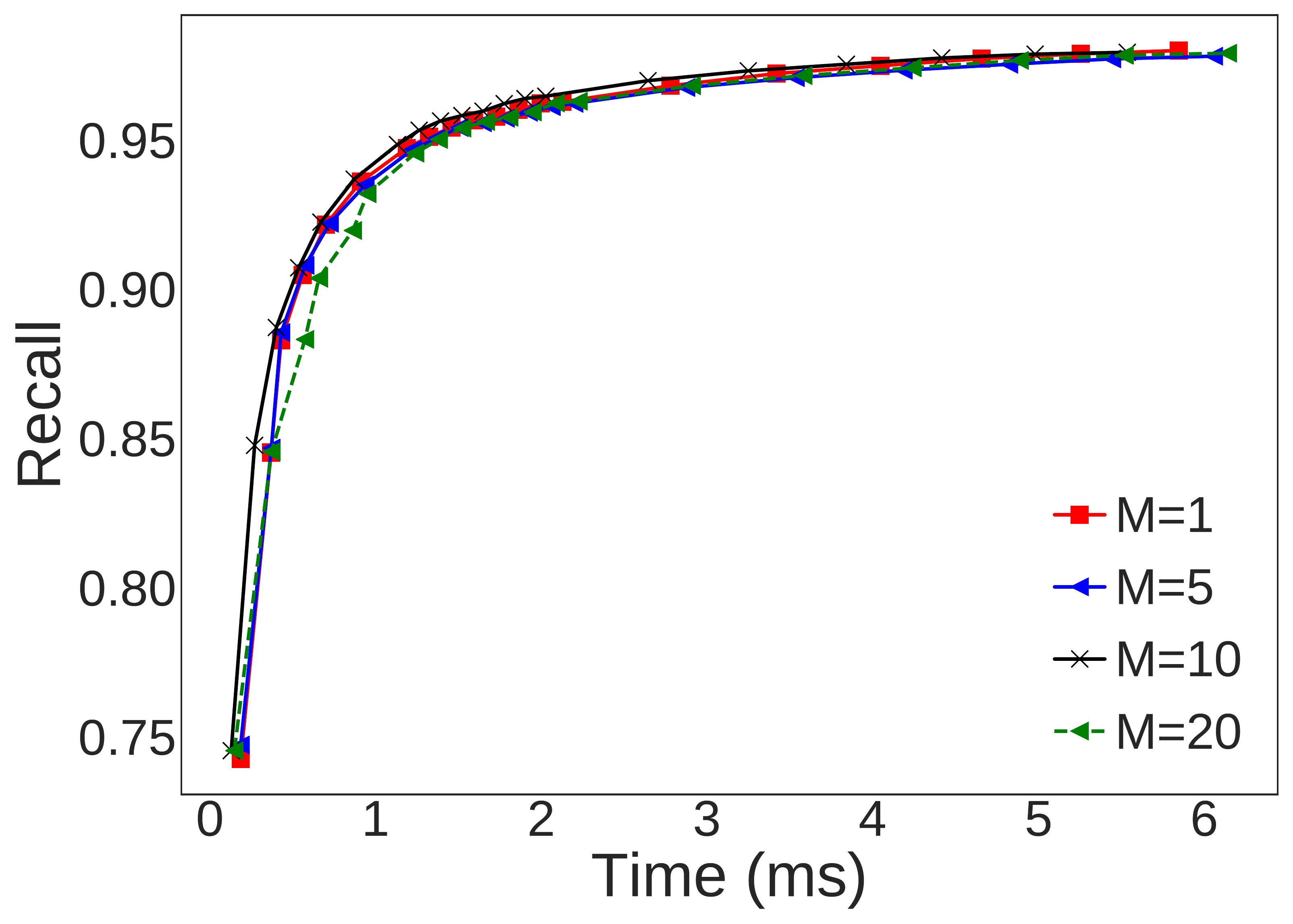}
		\subcaption{Changing $M$ with $l\!=\!10$}\label{fig:m}
	\end{minipage}
	\begin{minipage}[b]{0.23\textwidth}
		\includegraphics[width=\textwidth]{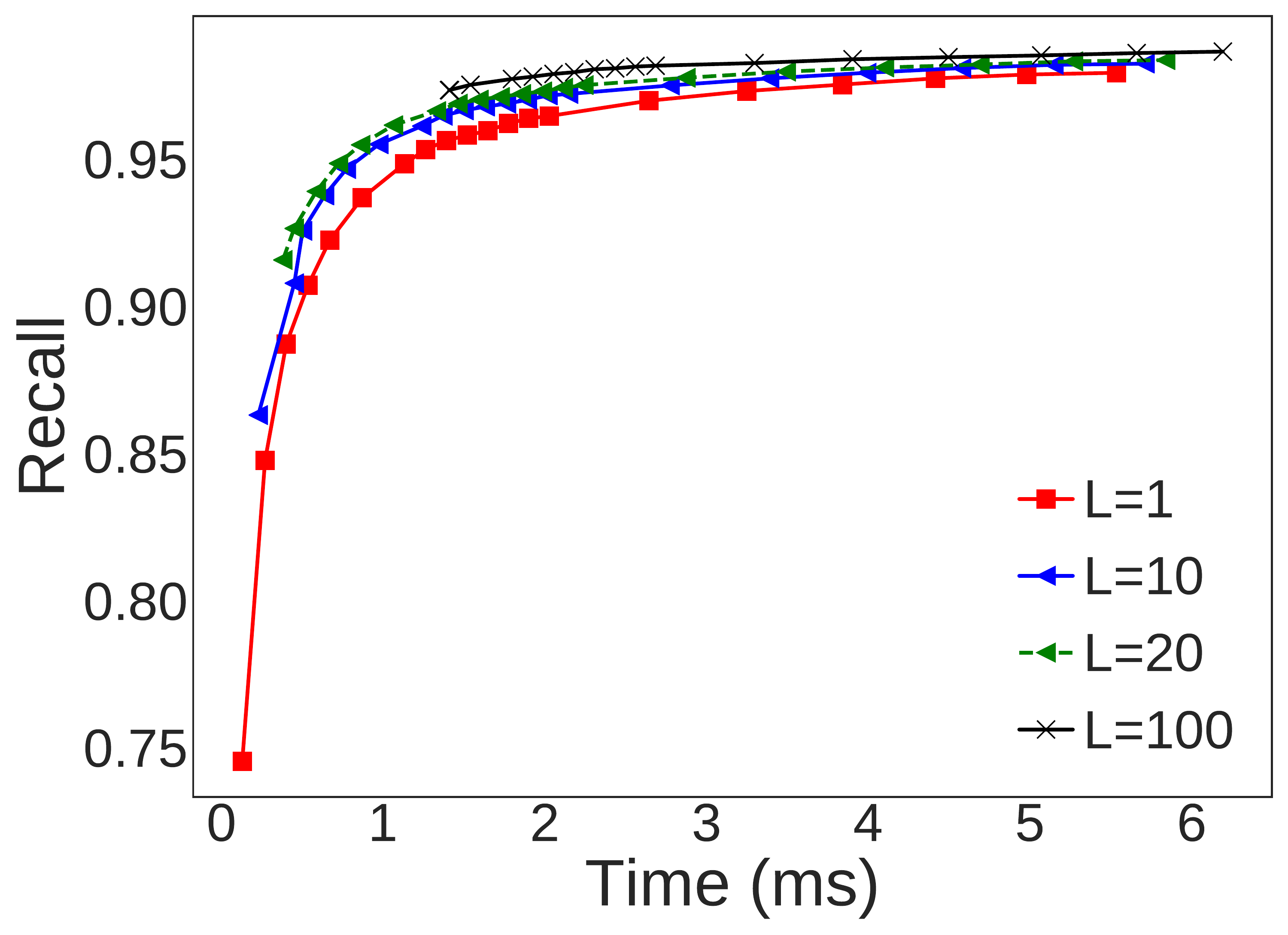}
		\subcaption{Changing $l$ with $M\!=\!10$}
		\label{fig:l}
	\end{minipage}
	\caption{Performance under different $M$ and $l$ on ImageNet} 
	\label{fig:parameter}
\end{figure}

We try to explain this phenomenon as follows. The norm bias in ip-NSW is more severe when the norm distribution is more skewed. Therefore, ip-NSW will compute inner product with more large norm items that do not have a good inner product with the query and hence its performance worsens. In contrast, ip-NSW+ collects the MIPS neighbors of the angular neighbors and these neighbors are shown to have a good inner product with the query in Theorem~\ref{theorem:ip}. The stable performance of ip-NSW indicates that it effectively avoids computing inner product with items having large norm but not likely to be the results of MIPS.  

We report the performance of ip-NSW+ when using different configurations of $M$ and $l$ for the angular graph $\mathcal{A}_s$ in Figure~\ref{fig:parameter}. Both $M$ and $l$ control the time spent on searching the angular neighbors, and setting a large value for them produces more accurate angular neighbors at the cost of using more time. The results show that the performance gap between different configurations is small and hence ip-NSW+ is robust to the choice of $M$ and $l$. The curves in Figure~\ref{fig:l} start at different time because the time spent on angular neighbor search varies a lot under different $l$ (but not the case for $M$), and ip-NSW+ produces MIPS results only after switching to the inner product graph. The results also show that using both too small values (e.g., $M\!=\!1$ or $l\!=\!1$, producing angular neighbors of poor quality) and too large values (e.g., $M\!=\!20$ or $l\!=\!100$, spending too much time for angular neighbor search) for the two parameters degrades performance. Therefore, we recommend to use the defualt setting with $M\!=\!10$ and $l\!=\!10$.

\section{Conclusions}\label{sec:conclusions}

In this paper, we identified an interesting phenomenon for the MIPS problem --- norm bias, which means that large norm items are much more likely to be the results of MIPS. We showed that ip-NSW achieves excellent performance for MIPS because it also has a strong norm bias, which means that the large norm items have large in-degrees in the ip-NSW graph and the majority of computation is conducted on them. We also proposed the ip-NSW+ algorithm, which avoids computation on large norm items that do not have a good inner product with the query. Experimental results show that ip-NSW+ significantly outperforms ip-NSW and is more robust to different data distributions. 

~\\
\textbf{Acknowledgments.} This work was supported by ITF 6904945, and GRF 14208318 \& 14222816, and the National Natural Science Foundation of China (NSFC) (Grant No. 61672552). 

\bibliography{3224.graphmips}
\bibliographystyle{aaai}

\newpage
\onecolumn
	\vbox{
	\hsize\textwidth
	\linewidth\hsize
	\vskip 0.1in
	\hrule height 4pt
	\vskip 0.25in
	\vskip -\parskip
	\centering
	{\LARGE\bf Supplementary Material for Understanding and Improving Proximity Graph based Maximum Inner Product Search\par}
	\vskip 0.29in
	\vskip -\parskip
	\hrule height 1pt                  
	\vskip 0.09in
}

\section{Proof of the Theorems}

In this part, we restate the two theorems in the main paper and provide proofs for them.

\begin{theorem}
	For two independent random vectors $x$ and $y$ in $\mathbb{R}^d$, the entries of $x$ are independent and $x_i\sim\mathcal{N}(0,\alpha)$ for $i=1,2,\cdots,d$ with $\alpha \ge 1$, the entries of $y$ are also independent and $y_i\sim\mathcal{N}(0,1)$ for $i=1,2,\cdots,d$. For a query $q\in\mathbb{R}^d$, we have $\mathbb{P}\lbrack q{\top}x\ge q{\top}y \mid  q{\top}x\ge 0, q{\top}y\ge 0 \rbrack = \frac{2}{\sqrt{\pi^2\alpha}} \int_{0}^{+\infty} e^{-\frac{a^2}{2\alpha}} \int_{0}^{a} e^{-\frac{b^2}{2}} \mathrm{d} b \mathrm{d} a$.  
	
	\begin{proof}
		We normalize the query to unit norm as $\bar{q}=\frac{q}{\Vert q \Vert}$ and define two random variables $a$ and $b$ with $a=\bar{q}^{\top}x$ and $b=\bar{q}^{\top}y$. $a$ and $b$ are independent and $a\sim\mathcal{N}(0,\alpha)$ and $b\sim\mathcal{N}(0,1)$. Therefore, their joint density function is $f(a, b)=f(a)f(b)=\frac{1}{\sqrt{2\pi\alpha}} e^{-\frac{a^2}{2\alpha}} \frac{1}{\sqrt{2\pi}} e^{-\frac{b^2}{2}}$. $\mathbb{P}\lbrack q{\top}x\ge q{\top}y \mid  q{\top}x\ge 0, q{\top}y\ge 0 \rbrack = \frac{\mathbb{P}\lbrack a\ge 0, b\ge 0, a\ge b\rbrack}{\mathbb{P}\lbrack a\ge 0, b\ge 0 \rbrack}=\frac{2}{\sqrt{\pi^2\alpha}} \int_{0}^{+\infty} e^{-\frac{a^2}{2\alpha}} \int_{0}^{a} e^{-\frac{b^2}{2}} \mathrm{d} b \mathrm{d} a$.  
	\end{proof}
\end{theorem} 

\begin{theorem}
	For two vectors $x$ and $y$ in $\mathbb{R}^d$ having an cosine similarity $\beta=\frac{x^{\top}y}{\Vert x \Vert \Vert y \Vert}$, a third vector $z\in\mathbb{R}^d$ and the entries of $z$ are independent and $z_i\sim\mathcal{N}(0,1)$ for $i=1,2,\cdots,d$, if $y^{\top}z=\gamma$, we have $x^{\top}z  \mid y^{\top}z=\gamma \sim\mathcal{N}(\frac{\gamma \beta \Vert x \Vert}{\Vert y \Vert},\Vert x \Vert^2(1-\beta^2))$. 
	\begin{proof}
		Define two random variables $a$ and $b$ with $a=x^{\top}z$ and $b=y^{\top}z$. $a\sim\mathcal{N}(0, \Vert x \Vert^2)$ and $b\sim\mathcal{N}(0, \Vert y \Vert^2)$, and $conv(a, b)=x^{\top}y=\beta \Vert x \Vert \Vert y \Vert $. Therefore, the covariance matrix $\mathbf{\Sigma}_{ab}=\begin{pmatrix} \Vert x \Vert^2 & \beta \Vert x \Vert \Vert y \Vert \\ \beta \Vert x \Vert \Vert y \Vert & \Vert y \Vert^2\end{pmatrix}$. The joint distribution of $a$ and $b$ can be expressed as $f(a, b)=\frac{1}{2\pi\sqrt{\vert \mathbf{\Sigma}_{ab} \vert}}e^{-\frac{1}{2}(a, b)^{\top}{\Sigma}_{ab}^{-1}(a, b)}$ with ${\Sigma}_{ab}^{-1}=\frac{1}{\Vert x \Vert^2 \Vert y \Vert^2 (1-\beta^2)}\begin{pmatrix} \Vert y \Vert^2 & -\beta \Vert x \Vert \Vert y \Vert \\ -\beta \Vert x \Vert \Vert y \Vert & \Vert x \Vert^2\end{pmatrix}$. Therefore, the density function of $a$ conditioned on $b=\gamma$ can be expressed as $f(a \mid b=\gamma)=\frac{f(a, b=\gamma)}{f_b(b=\gamma)}=\frac{1}{\sqrt{2\pi \Vert x \Vert^2 (1-\beta^2)}}e^{-\frac{(a-\frac{\gamma \beta \Vert x \Vert}{\Vert y \Vert})^2}{2\Vert x \Vert^2(1-\beta)^2}}$.	
	\end{proof}
	
\end{theorem}   

\section{Norm distributions of the ImageNet Variants}  

\begin{figure}[!h]	
	\centering 
	\begin{minipage}[b]{0.3\textwidth}
		\centering
		\includegraphics[width=\textwidth]{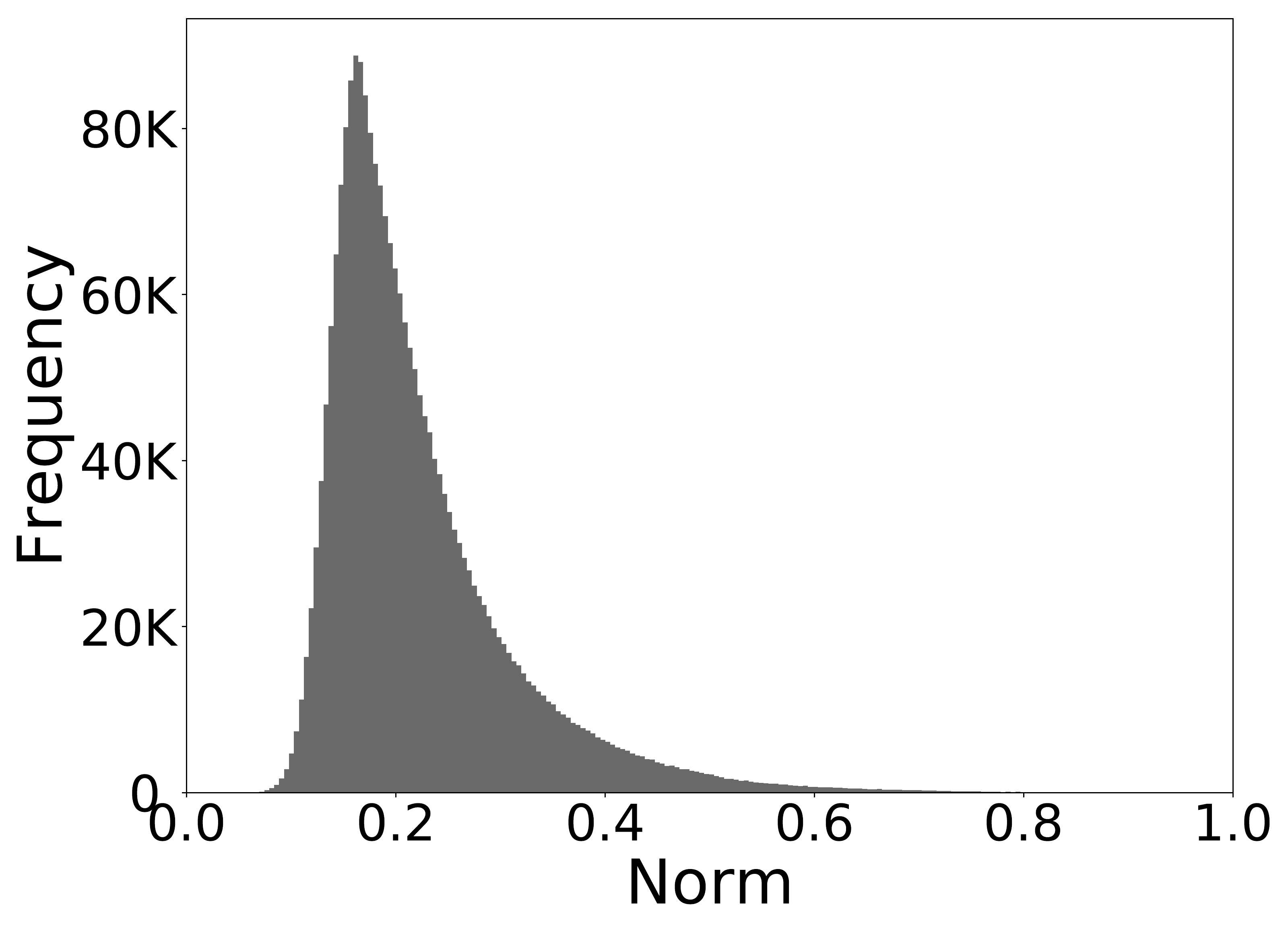}
		\subcaption{ImageNet}\label{fig:ImageNet}
	\end{minipage}
	\begin{minipage}[b]{0.3\textwidth}
		\centering
		\includegraphics[width=\textwidth]{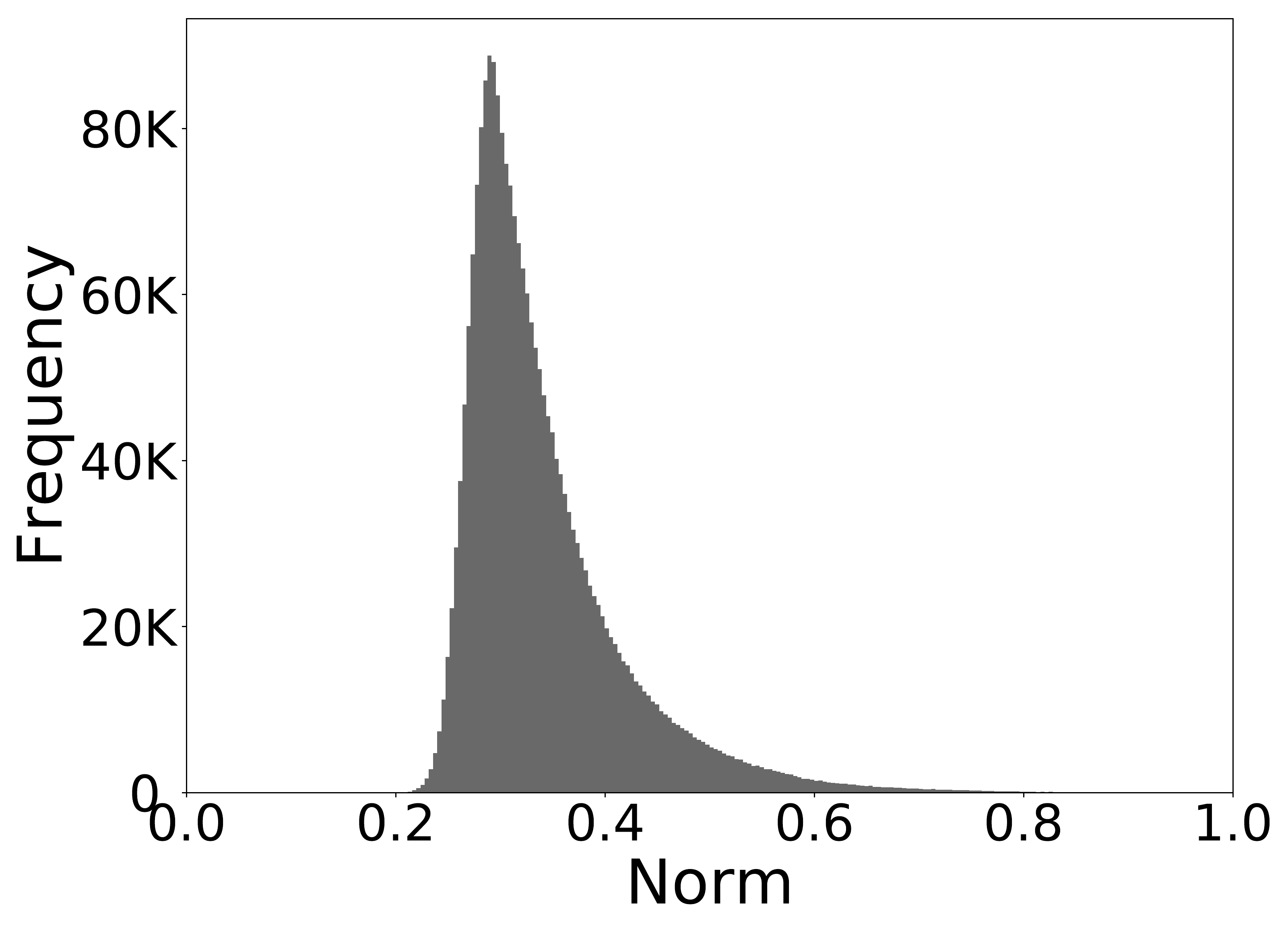}
		\subcaption{ImageNet-A}\label{fig:ImageNet-A}
	\end{minipage}
	\begin{minipage}[b]{0.3\textwidth}
		\centering
		\includegraphics[width=\textwidth]{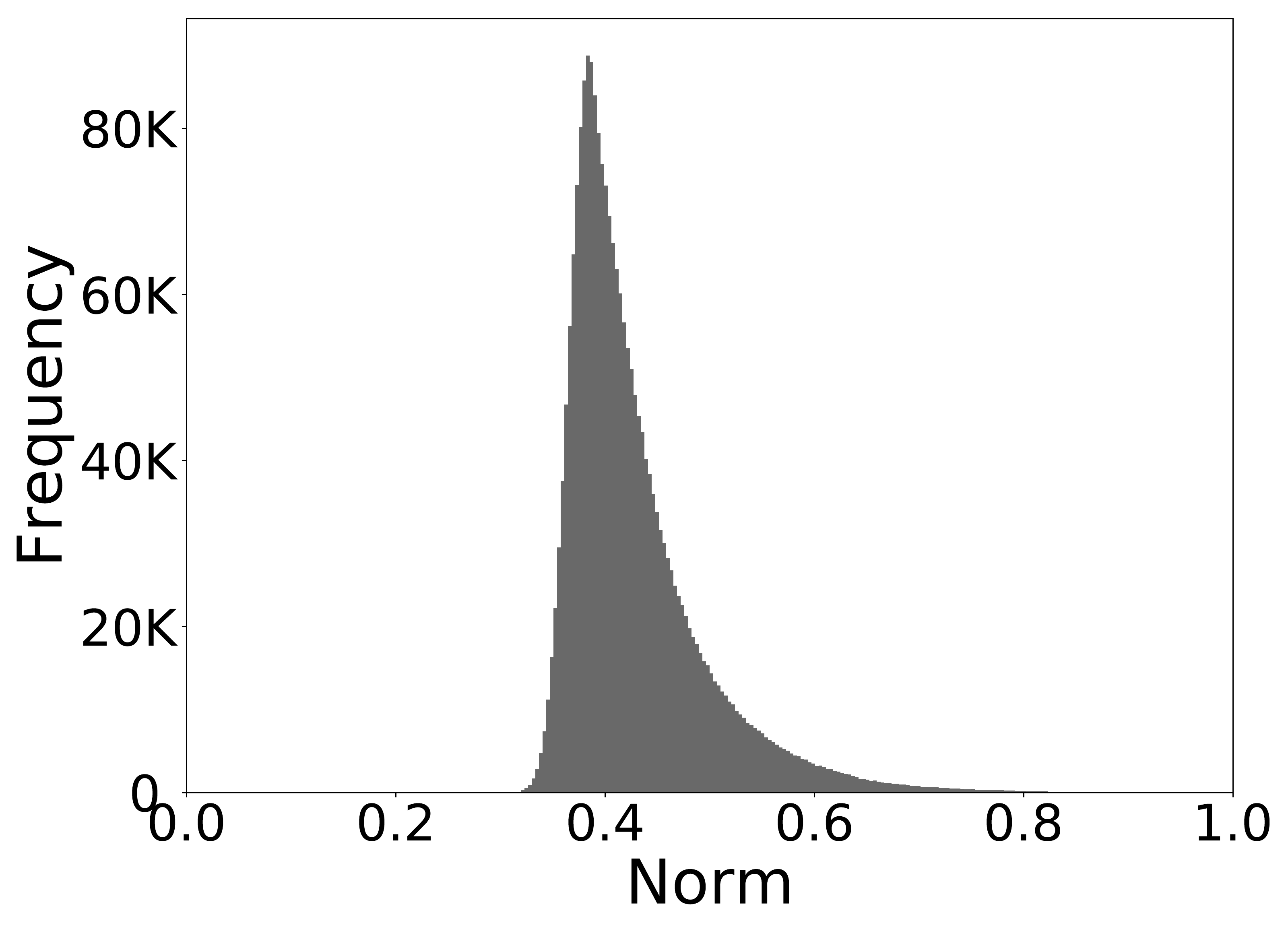}
		\subcaption{ImageNet-B}\label{fig:ImageNet-B}
	\end{minipage}
	\caption{Norm distributions of the ImageNet variants}
	\label{fig:imagenet variants}
\end{figure}

Figure~\ref{fig:imagenet variants} plot the norm distributions of the ImageNet variants in Section 5 of the main paper. ImageNet-A is created by adding 0.18 to the norm of each item while ImageNet-B adds 0.36 to the norm of each item. The tailing factors of ImageNet, ImageNet-A and ImageNet-B are 2.05, 1.55 and 1.37, respectively. Comparing to the average items, the top ranking items have the largest norm in the ImageNet dataset.

\section{Results of robustness on more datasets and an alternative scaling method}

\begin{figure*}[!h]
	\centering
	\includegraphics[width=0.32\textwidth]{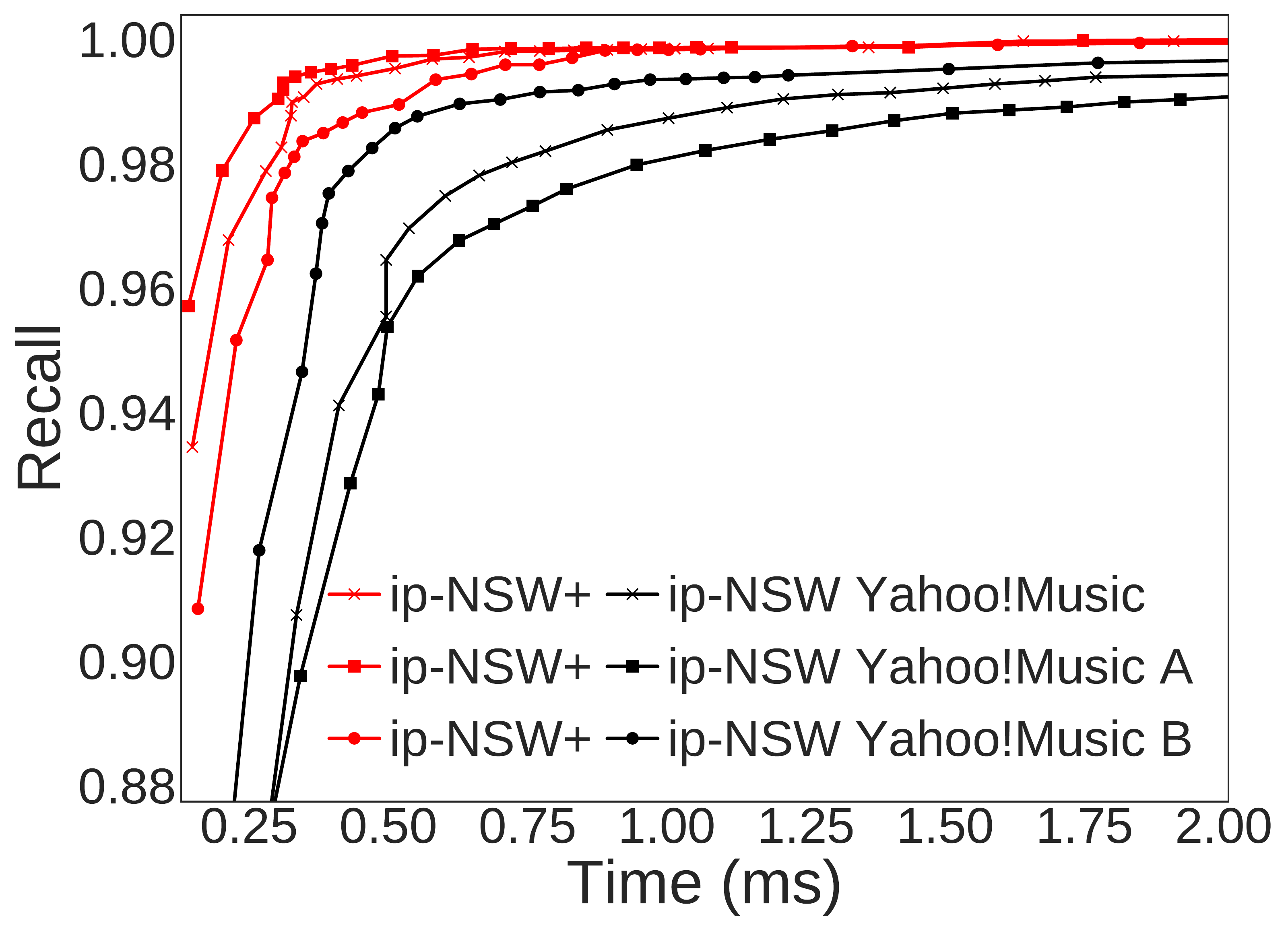}
	\includegraphics[width=0.32\textwidth]{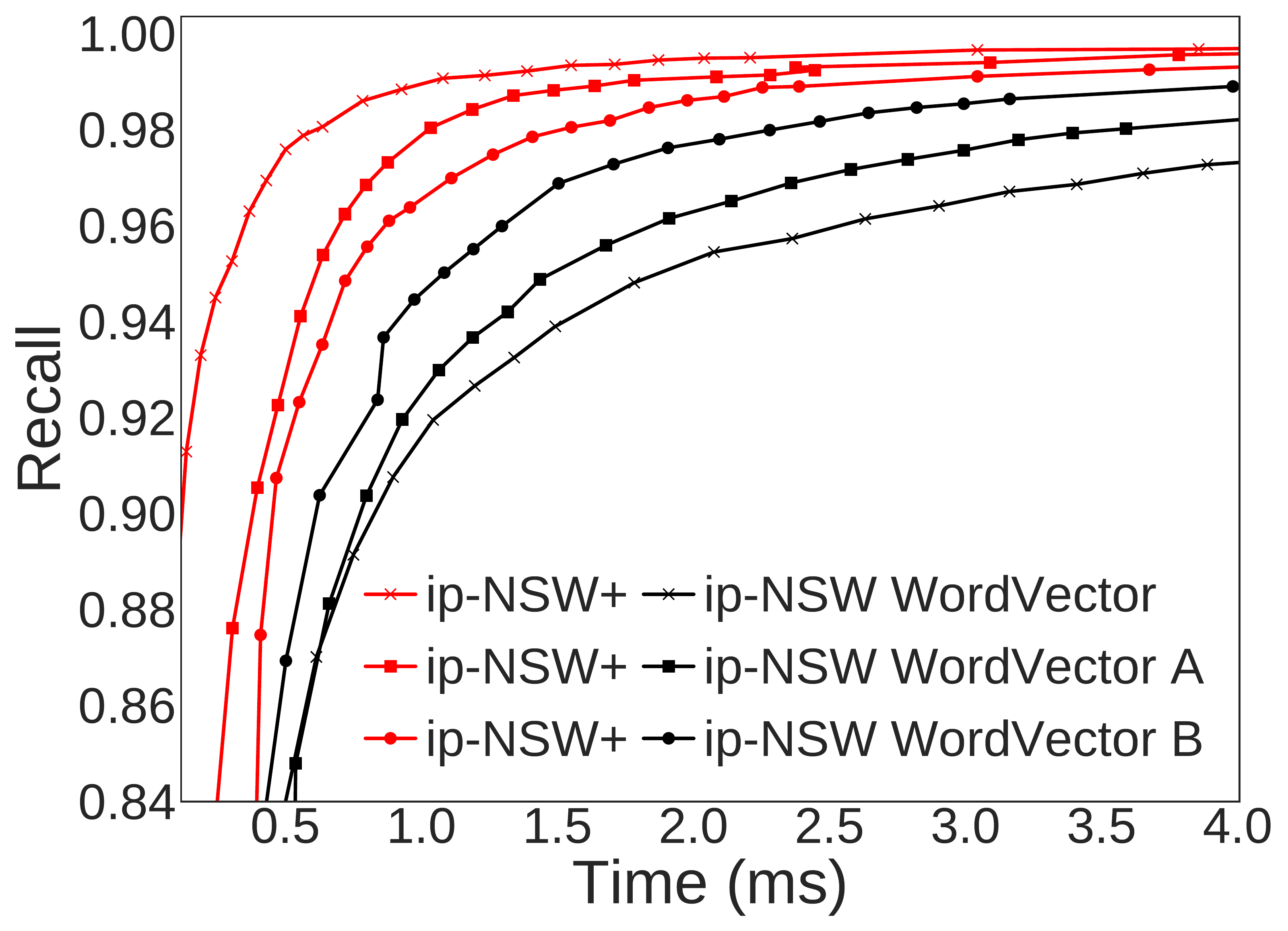}
	\caption{Performance comparison between ip-NSW and ip-NSW+ on variants of Yahoo!Music (left) and WordVector (right) (best viewed in colors)}
	\label{fig:scaling yahoomusict and wordvector}
\end{figure*}

To further verify the robustness of ip-NSW+ to data distribution, we modified the Yahoo!Music and WordVector dataset in a similar manner to the ImageNet dataset. Yahoo!Music-A is obtained by subtracting 0.2 from the norm of each item (which makes the tail of the norm distribution longer) while Yahoo!Music-B is obtained by adding 0.2 to the norm of each item (which makes the tail of the norm distribution shorter). WordVector-A and WordVector-B add 0.2 and 0.4 to the norm of each item, respectively. The performance results are shown in Figure~\ref{fig:scaling yahoomusict and wordvector}, which shows that ip-NSW+ has smaller gap between different variants of the same dataset.  

\begin{figure*}[!h]
	\centering
	\includegraphics[width=0.32\textwidth]{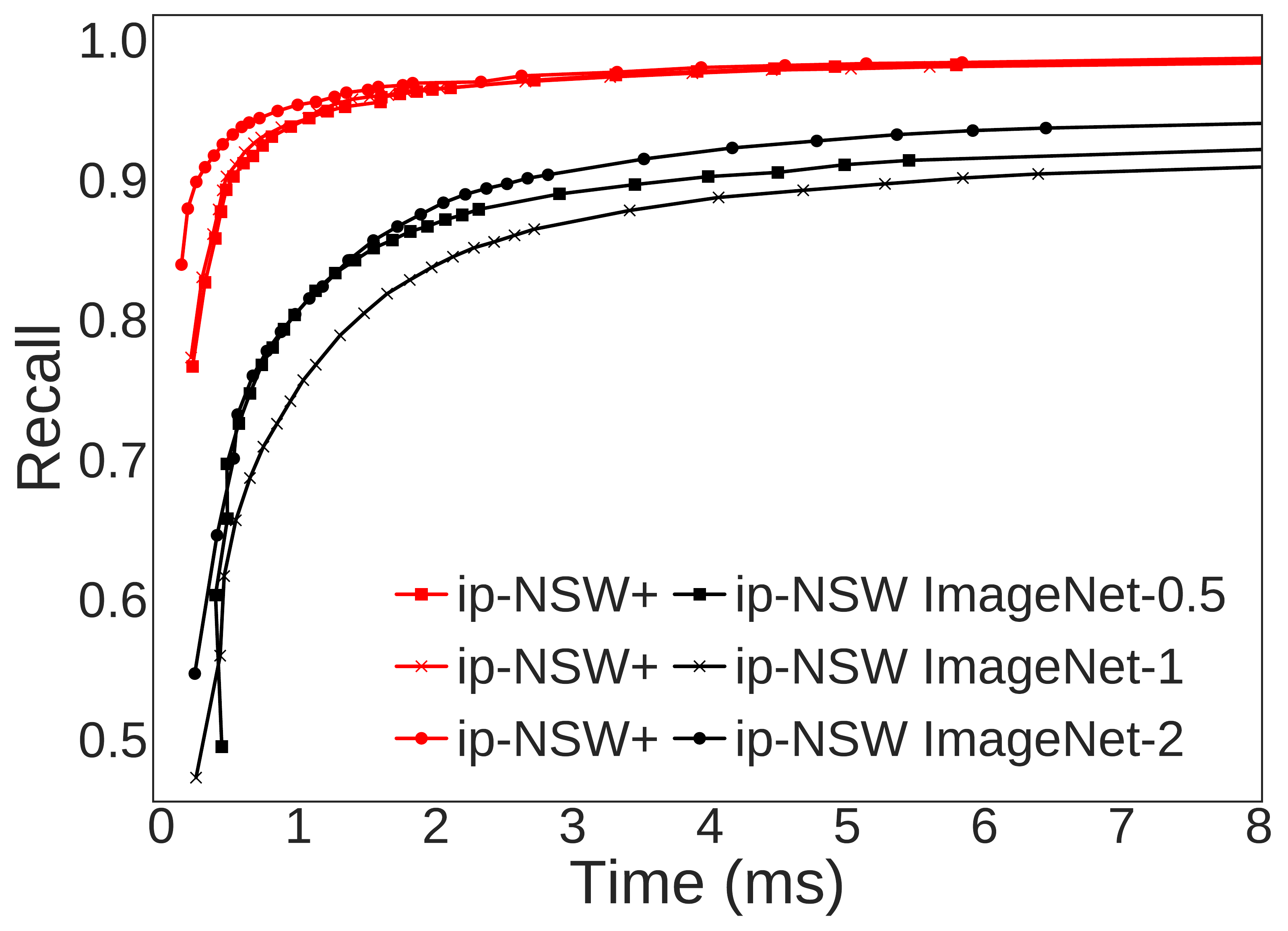}
	\caption{The performance of ip-NSW and ip-NSW+ on the ImageNet dataset under a new scaling method (best viewed in colors)}
	\label{fig:scaling}
\end{figure*}

We also experimented an alternative method to scale the norm of the items in the ImageNet dataset. Denote the scaling factor as $\beta$, the mode of the norm distribution as $t$ and the original norm of an item as $y$. The norm of an item is changed to $\tilde{y}=t-\beta (t-y)$ if $y\leq t$ and $\tilde{y}=t+\beta (y-t)$ otherwise. Intuitively, this scaling method controls the spread of the norm distribution and larger $\beta$ results in wider spread. We compare the performance of ip-NSW and ip-NSW+ under different scaling factors in Figure~\ref{fig:scaling}. The results show again that the performance of ip-NSW+ is more robust to data distribution. 


\section{Results for Simple-LSH}	
\begin{figure*}[!h]
	\centering
	\includegraphics[width=0.24\textwidth]{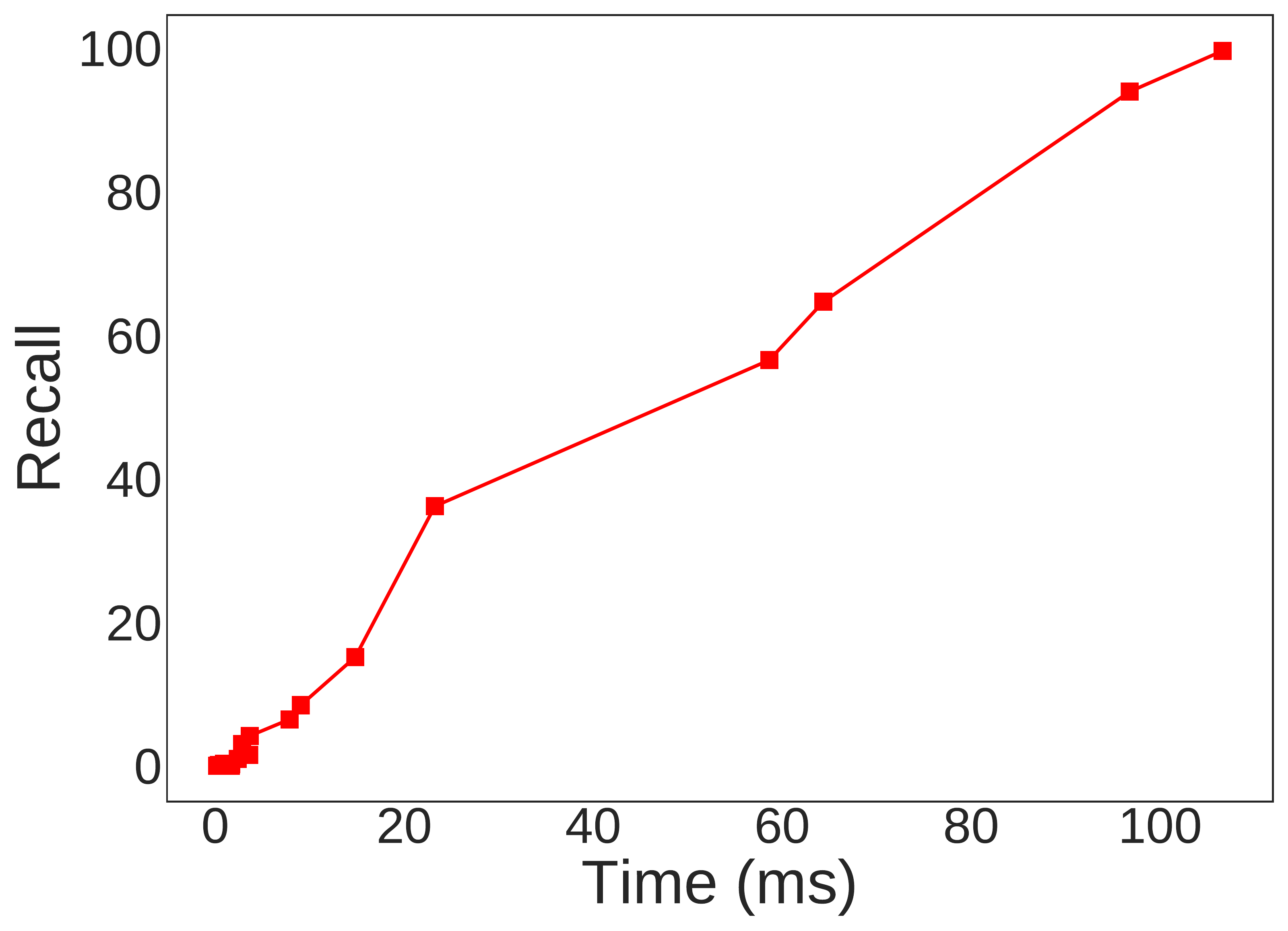}
	\includegraphics[width=0.24\textwidth]{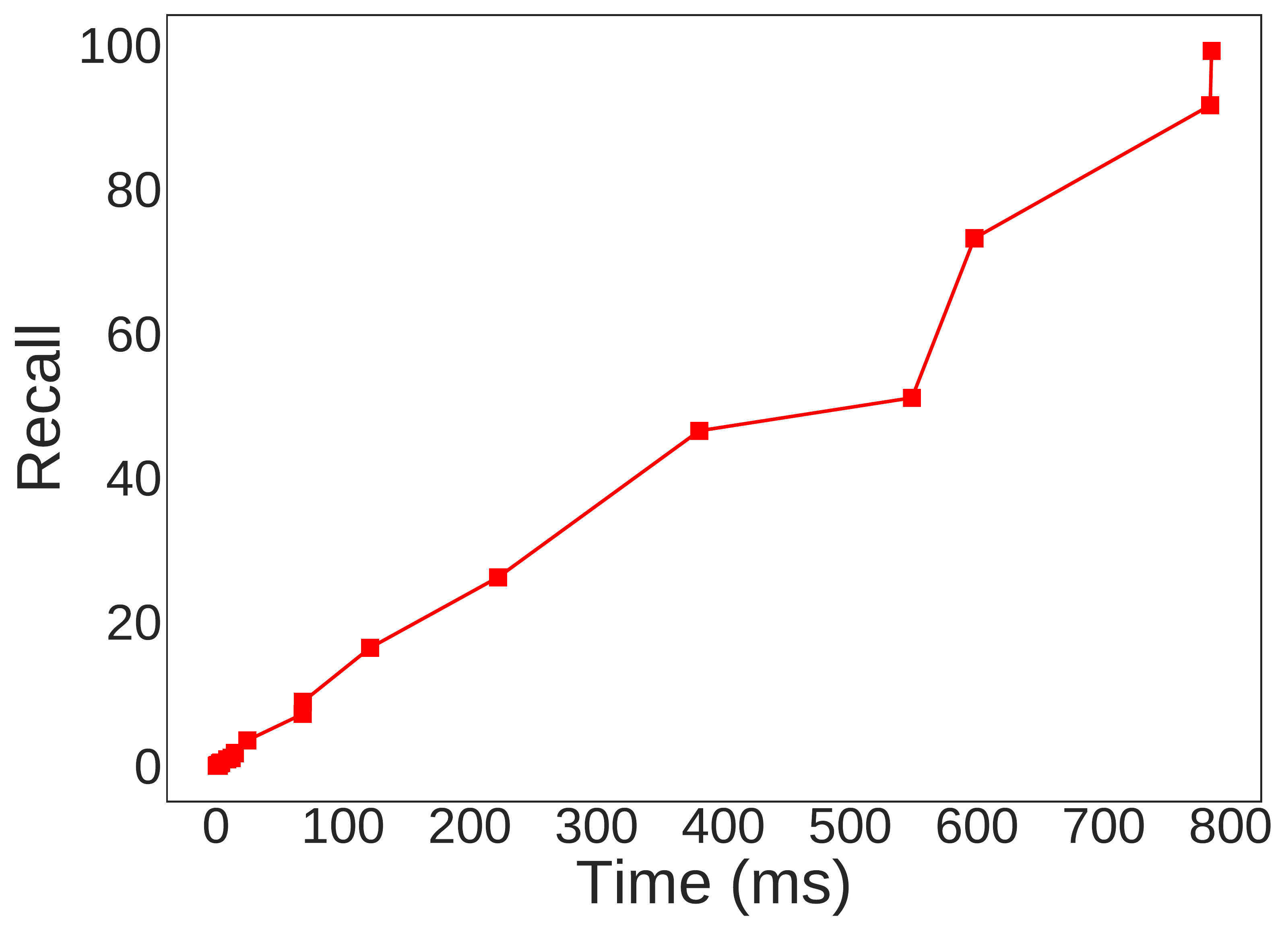}
	\includegraphics[width=0.24\textwidth]{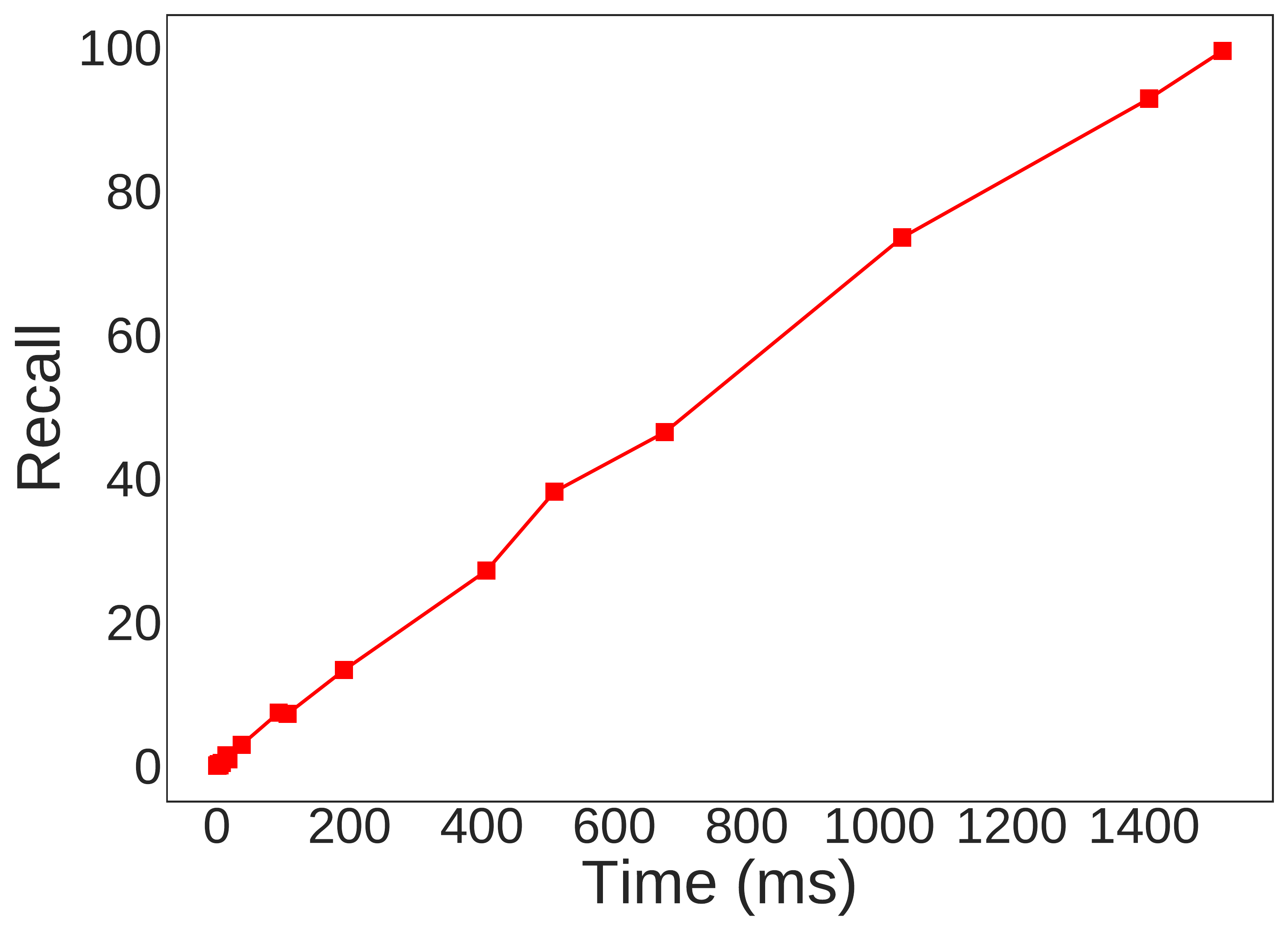}
	\includegraphics[width=0.24\textwidth]{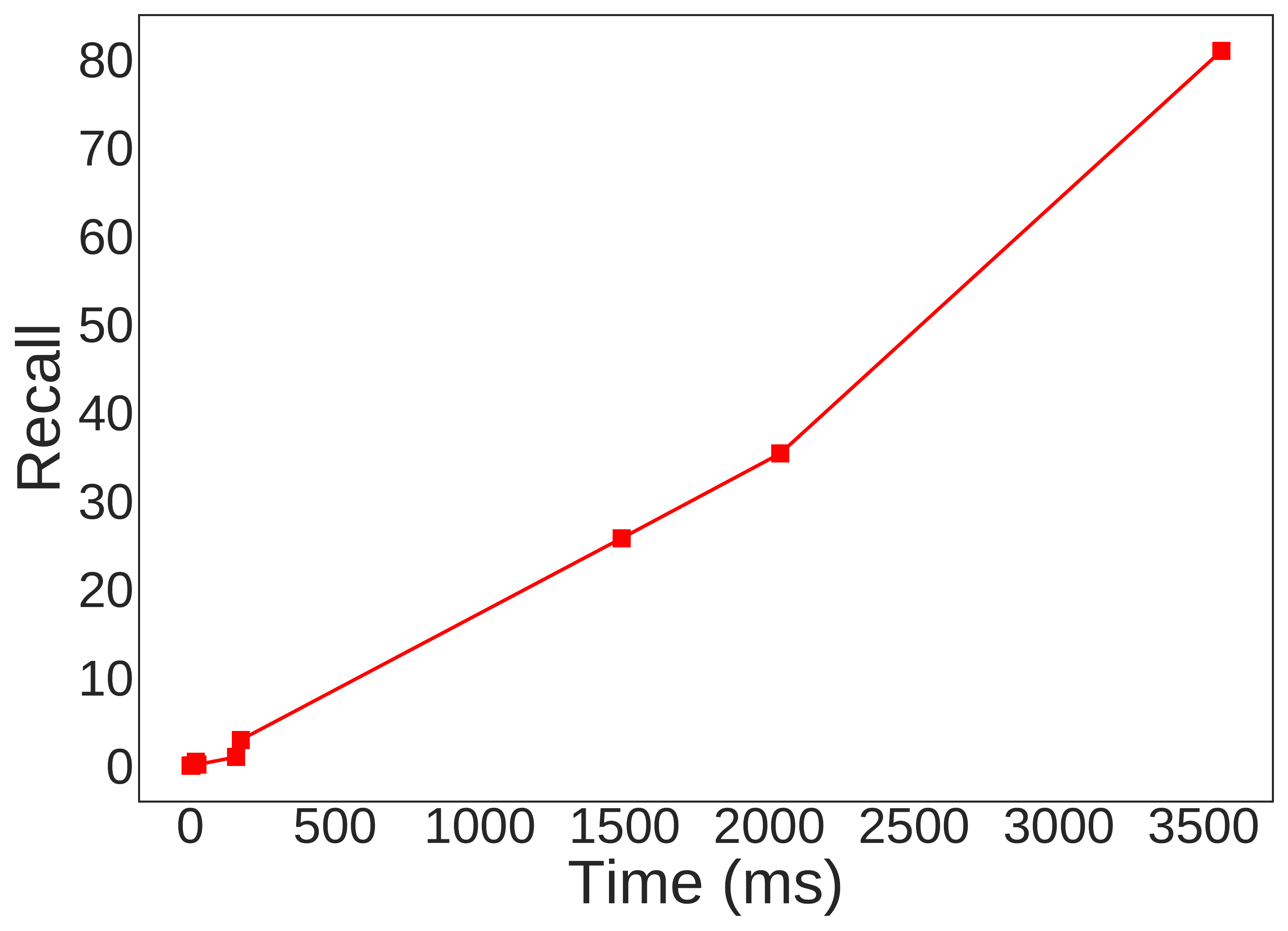}
	\caption{Recall-time performance of Simple-LSH, from left to right, the dataset is Yahoo!Music, WordVector, ImageNet and Tiny5M}
	\label{fig:simple-LSH}
\end{figure*}

\end{document}